\documentclass[aps, prc, tightenlines,superscriptaddress, letterpaper, amsmath, amssymb, preprintnumbers, floatfix, longbibliography, nofootinbib]{revtex4-2}
\usepackage[T1]{fontenc}
\usepackage{bm}
\usepackage{xspace}
\usepackage[dvipsnames]{xcolor}
\usepackage{graphicx}
\usepackage{tabularx}
\usepackage{enumitem}
\usepackage{braket}
\usepackage{dsfont}
\usepackage{placeins}
\usepackage[normalem]{ulem}
\usepackage{soul}

\newcolumntype{C}[1]{>{\centering\let\newline\\\arraybackslash\hspace{0pt}}m{#1}}
\newcolumntype{Y}{>{\centering\arraybackslash}X}

\usepackage{nccmath}
\usepackage{lipsum}
\usepackage{cancel}
\usepackage{bm}

\usepackage[hypertexnames=false]{hyperref}
\hypersetup{
    colorlinks=true,       
    linkcolor=blue,          
    citecolor=blue,        
    filecolor=blue,      
    urlcolor=blue           
}

\usepackage{mathtools}
\usepackage{orcidlink}

\definecolor{lavenderindigo}{rgb}{0.58, 0.34, 0.92}

\usepackage{stackengine}
\stackMath

\usepackage{setspace}

\begin{document}

\begin{figure}
\vskip -1.cm
\leftline{\includegraphics[width=0.15\textwidth]{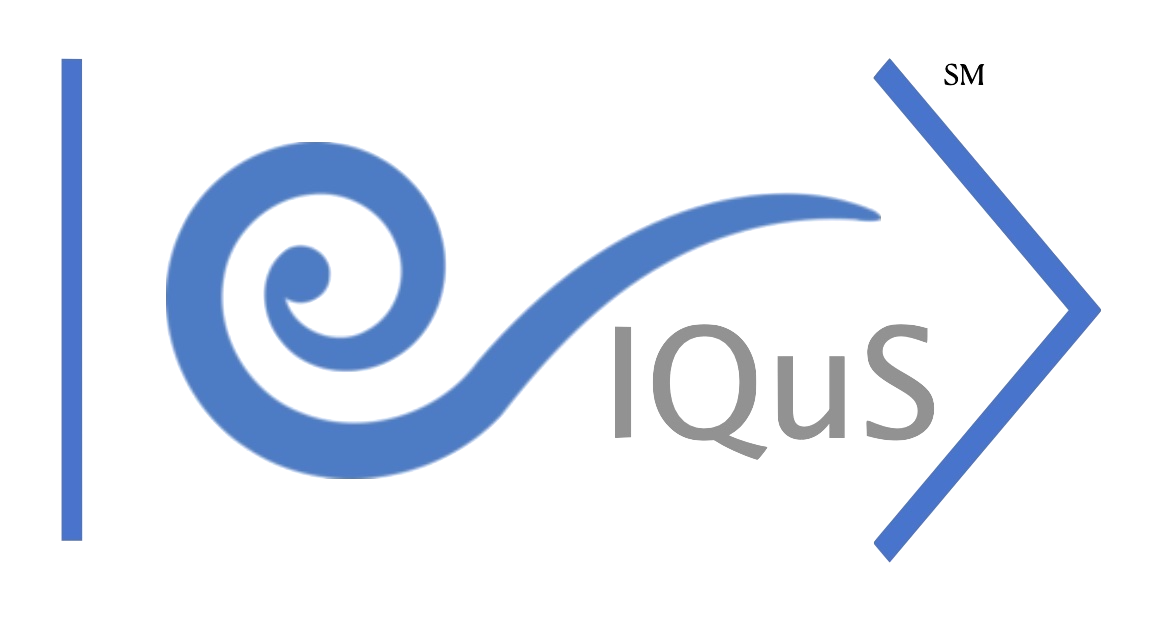}}
\vskip -0.5cm
\end{figure}

\title{Qu8its for Quantum Simulations of Lattice Quantum Chromodynamics}

\author{Marc Illa\,\orcidlink{0000-0003-3570-2849}}
\email{marcilla@uw.edu}
\affiliation{InQubator for Quantum Simulation (IQuS), Department of Physics, University of Washington, Seattle, WA 98195}

\author{Caroline E.~P.~Robin\,\orcidlink{0000-0001-5487-270X}}
\email{crobin@physik.uni-bielefeld.de}
\affiliation{Fakult\"at f\"ur Physik, Universit\"at Bielefeld, D-33615, Bielefeld, Germany}
\affiliation{GSI Helmholtzzentrum f\"ur Schwerionenforschung, Planckstra{\ss}e 1, 64291 Darmstadt, Germany}

\author{Martin J.~Savage\,\orcidlink{0000-0001-6502-7106}}
\email{mjs5@uw.edu}
\thanks{On leave from the Institute for Nuclear Theory.}
\affiliation{InQubator for Quantum Simulation (IQuS), Department of Physics, University of Washington, Seattle, WA 98195}

\preprint{IQuS@UW-21-074}
\date{\today}

\begin{abstract}
\noindent
We explore the utility of $d=8$ qudits, qu8its, for quantum simulations of the dynamics of 1+1D SU(3) lattice quantum chromodynamics, including a mapping for arbitrary number of flavors and lattice size and a re-organization of the Hamiltonian for efficient time-evolution.
Recent advances in parallel gate applications, along with the shorter application times of single-qudit operations compared with two-qudit operations, lead to significant projected advantages in quantum simulation fidelities and circuit depths using qu8its rather than qubits.
The number of two-qudit entangling gates required for time evolution using qu8its is found to be more than a factor of five fewer than for qubits.
We anticipate that the developments presented in this work will enable improved quantum simulations to be performed using emerging quantum hardware.
\end{abstract}

\maketitle
\newpage{}
\tableofcontents
\newpage{}

\section{Introduction}
\label{sec:intro}
\noindent
Quantum simulations of Standard Model physics are expected to provide results and insights 
about fundamental aspects of the structure and dynamics of matter
that are not possible with experiment or 
with classical computing alone~\cite{Banuls:2019bmf,Guan:2020bdl,Klco:2021lap,Alsing:2022ixr,Delgado:2022tpc,Bauer:2022hpo,Bauer:2023qgm,Beck:2023xhh,DiMeglio:2023nsa}.  
A major objective for high-energy physics and nuclear physics quantum simulations is to perform real-time simulations of non-equilibrium dynamics, such as of the low-viscosity quark-gluon liquid produced in heavy-ion collisions~\cite{Cohen:2021imf,Turro:2024pxu}, of the high-multiplicity events produced in proton-proton collisions~\cite{Bauer:2021gup}, and the creation of matter in the early Universe~\cite{Zhou:2021kdl}.
Progress toward such simulations is currently at early stages, in terms of the capabilities of quantum computers, of the sophistication of relevant quantum algorithms and workflows, and in understanding 
how to co-design quantum computers to best simulate quantum field theories.
A degree of focus is being placed on 1+1D systems, 
such as the Schwinger model, SU(3) quantum chromodynamics (QCD) 
and its SU(2) analog, and the Gross-Neveu model, 
in order to prepare for simulating QCD in 2+1D and 3+1D, 
with the ultimate goal of 
providing robust results (with a complete quantification of uncertainties)
for observables that cannot be assessed in the laboratory or with observation.

Most of the development so far has been centered around quantum computers utilizing qubits~\cite{Martinez:2016yna,Klco:2018kyo,Kokail:2018eiw,Lu:2018pjk,Klco:2019evd,Surace:2019dtp,Mil:2019pbt,Yang:2020yer,Bauer:2021gup,Zhou:2021kdl,Ciavarella:2021nmj,Atas:2021ext,ARahman:2021ktn,Mazzola:2021hma,deJong:2021wsd,Gong:2021bcp,Riechert:2021ink,Nguyen:2021hyk,Ciavarella:2021lel,Alsing:2022ixr,Illa:2022jqb,Mildenberger:2022jqr,Ciavarella:2022zhe,ARahman:2022tkr,Asaduzzaman:2022bpi,Mueller:2022xbg,Farrell:2022wyt,Atas:2022dqm,Farrell:2022vyh,Charles:2023zbl,Pomarico:2023png, PhysRevResearch.5.023010,zhang2023observation, Ciavarella:2023mfc,Borzenkova:2023xaf,Schuster:2023klj,Angelides:2023noe,Farrell:2023fgd,Chai:2023qpq,Farrell:2024fit,Kavaki:2024ijd,Turro:2024pxu,Davoudi:2024wyv,Ciavarella:2024fzw},
such as 
trapped-ion systems and superconducting qubit systems. 
However, in pursuit of hybrid qubit-qudit architectures~\cite{Meth:2023wzd,Zache:2023cfj}, 
further device capabilities, 
such as vibrational excitations~\cite{Mezzacapo:2012,Yang:2016hjn,Zhang:2016lyo,Davoudi:2021ney,Gonzalez-Cuadra:2023rex} or superconducting radio-frequency (SRF) cavities~\cite{Romanenko:2018nut,Ciavarella:2021nmj,Alam:2022crs,Roy:2024uro},
are being integrated into system functionalities.
Compared to qubits alone, quantum simulations using higher-dimensional qudits~\cite{Gottesman:1998se},\footnote{For a recent review of qudits, see, for example, Ref.~\cite{Wang:2020a}.} 
which can allow for ``better fit'' Hilbert spaces and reductions in the number of entangling gates~\cite{Illa:2023scc}, take us, in a sense, one more step away from classical computing.
With recent developments in quantum hardware, including results from qudit trapped-ion systems~\cite{Low:2019,Ringbauer:2021lhi,Low:2023dlg,Zalivako:2024bjm}, superconducting circuits~\cite{Blok:2020may,Seifert:2023ous,Nguyen:2023svc}, photonic systems~\cite{Chi:2022}, and nitrogen vacancy centers in diamond~\cite{Zhou:2023xnx}, 
and anticipating that such devices will become increasingly capable and available (in particular, trapped-ion systems with $d>10$~\cite{Low:2023dlg}),
a more general consideration of how one can utilize qudits in simulations of Standard Model quantum field theories is timely.
Examples of such applications can be found in Refs.~\cite{Ciavarella:2021nmj,Gustafson:2021qbt,Gustafson:2022xlj,Gonzalez-Cuadra:2022hxt,Gustafson:2022xdt,Gustafson:2023swx,Zache:2023cfj,Popov:2023xft,Meth:2023wzd,Calajo:2024qrc,Carena:2024dzu}, 
where the larger Hilbert space of qudits is used to describe the gauge fields of 
(non-)Abelian lattice gauge theories, or in Refs.~\cite{Calixto_2021,Illa:2023scc}, 
where nuclear many-body systems naturally map to qudits.

Quantum simulations of 1+1D SU(2)~\cite{Klco:2019evd,ARahman:2021ktn,ARahman:2022tkr,Atas:2021ext,Kavaki:2024ijd,Turro:2024pxu} and SU(3)~\cite{Ciavarella:2021nmj,Ciavarella:2021lel,Illa:2022jqb,Farrell:2022wyt,Atas:2022dqm,Farrell:2022vyh,Ciavarella:2023mfc,Ciavarella:2024fzw} lattice gauge theories have recently been performed.
Those that included matter fields have used the Kogut-Susskind (KS) staggered discretization\footnote{Different discretization approaches, such as Wilson fermions, are also being investigated~\cite{Zache:2018jbt,Mathis:2020fuo,Mazzola:2021hma,Bermudez:2023nve,Hayata:2023skf}.} 
of the quark fields~\cite{Kogut:1974ag,Banks:1975gq}, 
and worked in axial gauge to utilize Gauss's law to uniquely define the gauge fields~\cite{Sala:2018dui,Farrell:2022wyt}, 
enabling their contributions to be included by non-local all-to-all interactions. 
In this mapping, one color of one flavor of quark is mapped to two qubits, one describing the occupation of that quark and one of the corresponding  anti-quark.
For example, to describe one site of two-flavor ($N_f=2$) SU(3) QCD requires 12 qubits.\footnote{The lepton fields have also been included, in simulating the $\beta$-decay of a baryon~\cite{Farrell:2022vyh}, requiring four additional qubits per lepton generation.} 
Quantum circuits for preparing the ground state and implementing Trotterized time evolution have been identified and the quantum resources established (providing an upper bound)~\cite{Farrell:2022wyt}.

In this work, we explore the utility of qudits with $d=8$, which we denote as qu8its,\footnote{Which we suggest is pronounced {\it q-huits}.}
in simulating 1+1D lattice QCD using the KS discretization.
This is motivated,
in part,
by our demonstration of the utility of using qu5its in quantum simulations of 
multi-fermion (nucleon) systems with pairing interactions~\cite{Illa:2023scc} 
and an 
underlying
SO(5) symmetry.  
The 8 states associated with a single quark flavor mapped to three qubits
can be mapped to the states of a single qu8it, and the 8 states associated with single anti-quark mapped to three qubits can be mapped to another 
qu8it (which we loosely denote as an anti-qu8it).
While it does not seem to be an immediate gain by counting the number of states, 
the advantage is in the number of entangling gates required to implement time evolution, and in the number of units in a quantum register.
Analogous to qubit operations, 
the single-qudit gate operations are significantly faster and of higher fidelity than 
the two-qudit operations.\footnote{For example, in IonQ's current flagship trapped-ion qubit quantum computer, single-qubit operations require $\sim 100\; \mu s$ 
while two-qubit operations require $\sim 600 \;\mu s$~\cite{IonQaria}.
}
Consequently, the factor of $\gtrsim 5$ reduction in the number of two-qudit gates 
and the factor of 3 reduction in the number of units in the register that we find in mapping to qu8its, 
suggest
that quantum computers with qu8its are likely to provide enhanced capabilities for simulating 
non-Abelian lattice gauge theories.

\section{The Kogut-Susskind Hamiltonian for QCD and Mapping to Qubits}
\label{sec:KSqubits}
\noindent
The 1+1D SU(3) KS Hamiltonian~\cite{Kogut:1974ag,Banks:1975gq} with $N_f$ flavors formulated in $A^{(a)}_x = 0$ gauge~\cite{Sala:2018dui,Farrell:2022wyt} takes the form
\begin{align}
    H 
    = & 
    \sum_{f}\left[ 
        \frac{1}{2} \sum_{n=0}^{2L-2} \left ( \phi_n^{(f)}{}^\dagger \phi_{n+1}^{(f)}
        \ +\ {\rm h.c.} \right ) 
    \: + \: 
    m_f \sum_{n=0}^{2L-1} (-1)^{n} \phi_n^{(f)}{}^\dagger \phi_n^{(f)} 
    \right]
    \: + \: 
    \frac{g^2}{2}
    \sum_{n=0}^{2L-2} 
    \sum_{a=1}^{8}
    \left ( \sum_{m \leq n} Q^{(a)}_m \right ) ^2    
    \ ,
    \label{eq:GFHam}
\end{align}
where $\phi_n^{(f)}$ correspond to annihilation operators for fermions of flavor $f$.
They are color triplets, with their color indices suppressed in Eq.~(\ref{eq:GFHam}).
The color-charge operators on each lattice site are the sum of contributions 
from each flavor. For example, for $N_f=2$ (up and down quarks), the color-charge operators are
\begin{equation}
    Q^{(a)}_m \ =\ 
    \phi^{(u) \dagger}_m T^a \phi_m^{(u)}\ +\ 
    \phi^{(d) \dagger}_m T^a \phi_m^{(d)}
    \ ,
    \label{eq:SU3charges}
\end{equation}
where the generators of SU(3), $T^a$,  are given in App.~\ref{ap:GM}.
With open boundary conditions (OBC) and vanishing fields at spatial infinity, corresponding to vanishing net color charge on the lattice (enforced by additional terms in the Hamiltonian~\cite{Farrell:2022wyt}), 
Gauss's law is sufficient to determine the chromo-electric field at all lattice sites,
\begin{equation}
   {\bf E}^{(a)}_n  = \sum_{m\leq n} Q^{(a)}_m 
   \ .
\end{equation}

\begin{figure}[!ht]
    \centering
    \includegraphics[width=0.95\textwidth]{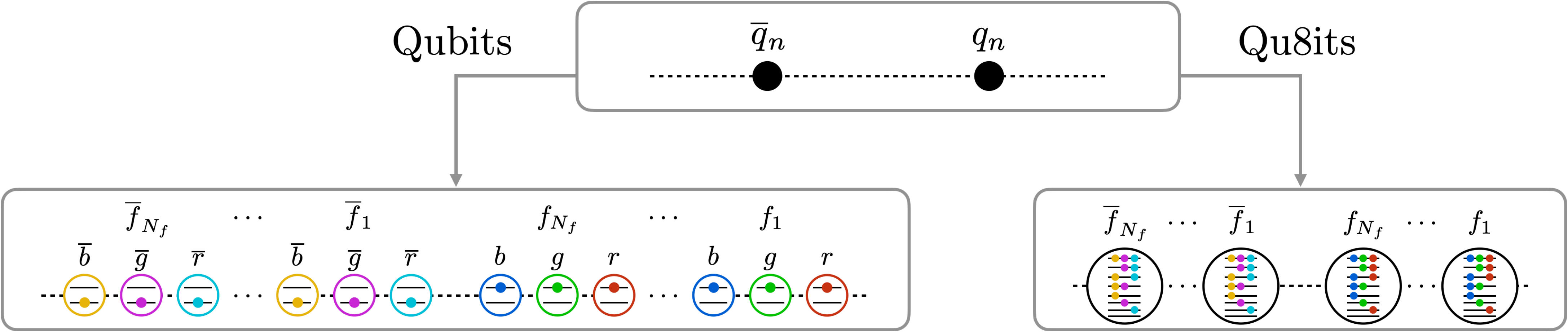}
    \caption{
    Mapping  QCD with $N_f$ quark flavors onto a lattice of qubits (left) or qu8its (right) 
    describing a spatial site. 
    Kogut-Susskind (staggered) fermions are used for the quark fields, with (anti)quarks on (odd) even sites.
    Using qubits, color and flavor degrees of freedom of each quark and antiquark site are distributed over six qubits with a JW mapping.
    Using qu8its, with the quark (and anti-quark) degrees of freedom being mapped to the internal states, only two qu8its are required per each quark flavor.}
    \label{fig:2flavLayout}
\end{figure}

There are a number of ways that this system,
with the Hamiltonian given in Eq.~\eqref{eq:GFHam}, can be mapped
onto qubit registers.
In our previous works~\cite{Farrell:2022wyt,Farrell:2022vyh},
the KS Hamiltonian for an arbitrary number of colors $N_c$ and flavors $N_f$
was mapped to qubits using the Jordan-Wigner (JW) transformation~\cite{Jordan:1928wi}. For the $N_c=3$ and $N_f=2$ case,
each staggered site requires six qubits, 
with ordering $d_b, d_g, d_r, u_b, u_g, u_r$,
and the antiquarks associated with the same spatial site adjacent with ordering 
$\overline{d}_b, \overline{d}_g, \overline{d}_r, \overline{u}_b, \overline{u}_g, \overline{u}_r$.
This is shown in the left panel of Fig.~\ref{fig:2flavLayout}.
The resulting JW-mapped Hamiltonian is the sum of three terms~\cite{Farrell:2022wyt,Farrell:2022vyh}, neglecting the possible presence of chemical potentials,
\begin{eqnarray}
    \label{eq:H2flav}
    H & = & \ H_{kin}\ +\ H_m\ +\ H_{el} \ , \nonumber \\
    H_{kin} 
    & = & \ -\frac{1}{2} \sum_{n=0}^{2L-2} \sum_{f=0}^{1} \sum_{c=0}^{2} \left[ \sigma^+_{6n+3f+c} \left ( \bigotimes_{i=1}^{5}\sigma^z_{6n+3f+c+i} \right )\sigma^-_{6(n+1)+3f+c} +\rm{h.c.} \right]\ ,
        \nonumber\\ 
    H_m & = & \ \frac{1}{2} \sum_{n=0}^{2L-1} \sum_{f=0}^{1} \sum_{c=0}^{2} m_f\left[ (-1)^{n} \sigma_{6n + 3f + c}^z + 1\right]\ ,
        \nonumber \\ 
    H_{el} & = & \ \frac{g^2}{2} \sum_{n=0}^{2L-2}(2L-1-n)\left( \sum_{f=0}^{1} Q_{n,f}^{(a)} \, Q_{n,f}^{(a)} + 
        2 Q_{n,0}^{(a)} \, Q_{n,1}^{(a)}
         \right)  
         + g^2 \sum_{n=0}^{2L-3} \sum_{m=n+1}^{2L-2}(2L-1-m) \sum_{f=0}^1 \sum_{f'=0}^1 Q_{n,f}^{(a)} 
         \, Q_{m,f'}^{(a)} 
         \ ,
\end{eqnarray}
where repeated adjoint color indices $(a)$ are summed over,
the flavor indices $f=\{0,1\}$ correspond to $u$- and $d$-quark flavors, 
and $\sigma^\pm = (\sigma^x \pm i \sigma^y)/2$.
Products of charges, given in terms of spin operators, are given in 
Refs.~\cite{Farrell:2022wyt,Farrell:2022vyh},
\begin{align}
    Q_{n,f}^{(a)} \, Q_{n,f}^{(a)} \ = & \ \frac{1}{3}(3 - \sigma^z_{6n+3f} \sigma^z_{6n+3f+1} - \sigma^z_{6n+3f} \sigma^z_{6n+3f+2} - \sigma^z_{6n+3f+1} \sigma^z_{6n+3f+2}) \ ,  \nonumber \\[4pt]
    Q_{n,f}^{(a)} \, Q_{m,f'}^{(a)} \ = & \ \frac{1}{4}\bigg [2\big (\sigma^+_{6n+3f}\sigma^-_{6n+3f+1}\sigma^-_{6m+3f'}\sigma^+_{6m+3f'+1} + \sigma^+_{6n+3f}\sigma^z_{6n+3f+1}\sigma^-_{6n+3f+2}\sigma^-_{6m+3f'}\sigma^z_{6m+3f'+1}\sigma^+_{6m+3f'+2} \nonumber \\[4pt]
    &+\sigma^+_{6n+3f+1}\sigma^-_{6n+3f+2}\sigma^-_{6m+3f'+1}\sigma^+_{6m+3f'+2} + { \rm h.c.}\big ) + \frac{1}{6}\sum_{c=0}^{2} \sum_{c'=0}^2( 3 \delta_{c c'} - 1 ) \sigma^z_{6n+3f+c}\sigma^z_{6m+3f'+c'} \bigg ] \ .
    \label{eq:QnfQmfp}
\end{align}
A constant has been added to $H_m$ so that all basis states contribute a positive mass. 

The time-evolution operator corresponding to this Hamiltonian can be implemented by sequences of unitary operators. These Trotterized quantum circuits, and hence the 
associated gate counts per Trotter step, 
have been determined in Refs.~\cite{Farrell:2022wyt,Farrell:2022vyh}.

\section{Qu8its for QCD}
\label{sec:EandO}
The main drivers for considering mapping the KS Hamiltonian to qudits with $d=8$, as discussed above, is to reduce the number of two-qudit entangling gates from the number required for qubits.
The single-component fermion formulation (per color state) that defines the staggering of the quark fields
draws comparisons with the constructions used in describing nuclear many-body systems.
We recently showed the utility of using $d=5$ qudits (qu5its) to describe spin-paired multi-nucleon systems in the context of the Agassi model~\cite{Illa:2023scc}. We make use of this analogy in mapping the quark (and anti-quark) fields to qu8its, as displayed in the right panel of Fig.~\ref{fig:2flavLayout}.

\subsection{Mapping Quarks and Anti-Quarks to Qu8its}
\label{sec:qs}
\noindent
For a single flavor quark staggered site and using the JW mapping to three qubits,
the qubits define the occupation of the three colors
$q_r, q_g, q_b$ throughout a quantum simulation.  
Satisfying fermion anti-commutation relations,
the mapping of color occupations to qu8its can be chosen to be,
\begin{eqnarray}
    \big\{  |{\rm qu8it}\rangle  \big\}
    \ & = & \
    \big\{ 
    |\Omega\rangle \ , \ 
    |q_r\rangle\ , \ 
    |q_g\rangle\ , \ 
    |q_b\rangle\ , \ 
    |q_g q_b\rangle\ , \ 
    -|q_r q_b\rangle\ , \ 
    |q_r q_g\rangle\ , \ 
    |q_r q_g q_b\rangle
    \big\}
    \nonumber\\
    \ & = & \
    \big\{ 
    |\Omega\rangle \ , \ 
    \hat{c}_{r}^\dagger|\Omega\rangle\ , \ 
    \hat{c}_{g}^\dagger|\Omega\rangle\ , \ 
    \hat{c}_{b}^\dagger|\Omega\rangle\ , \ 
    \hat{c}_{g}^\dagger \hat{c}_{b}^\dagger|\Omega\rangle\ , \ 
    -\hat{c}_{r}^\dagger \hat{c}_{b}^\dagger|\Omega\rangle\ , \ 
    \hat{c}_{r}^\dagger \hat{c}_{g}^\dagger|\Omega\rangle\ , \ 
    \hat{c}_{r}^\dagger \hat{c}_{g}^\dagger \hat{c}_{b}^\dagger|\Omega\rangle  
    \big\}
    \nonumber\\
    \ & = & \
    \big\{ 
    |1\rangle \ , \ 
    |2\rangle\ , \ 
    |3\rangle\ , \ 
    |4\rangle\ , \ 
    |5\rangle\ , \ 
    |6\rangle\ , \ 
    |7\rangle\ , \ 
    |8\rangle
    \big\} 
    \ ,
    \label{eq:states}
\end{eqnarray}
where the fermionic vacuum state is $|\Omega\rangle$ and the 
$\hat{c}_{\alpha}$ operators are elements of the $\phi_j^{(f)}$ defined below Eq.~\eqref{eq:GFHam}.\footnote{The sign in the definition of $|6\rangle$ results from the $\epsilon^{132}$ in constructing a $\overline{\bf 3}$ from ${\bf 3}\otimes {\bf 3}$.}
At first, the group structure might be a little confusing, for instance the reason as to why there is only one state associated with the three quarks in the maximally occupied state ($|8\rangle$).
Given that quarks reside in the fundamental representation of SU(3), ${\bf 3}$, products of two and three quarks give rise to the following irreducible representations (irreps):
${\bf 3}\otimes {\bf 3}  =  {\bf 6} \oplus \overline{\bf 3}$ 
and ${\bf 3}\otimes {\bf 3} \otimes {\bf 3} =  {\bf 10} \oplus {\bf 8}\oplus {\bf 8} \oplus {\bf 1}$.
For spinless fermions at the same lattice site, only the total antisymmetric irreps
are allowed; therefore,
the irreps of the mapping in Eq.~\eqref{eq:states} are constrained to be
$\{{\bf 1}, {\bf 3}, {\bf 3}, {\bf 3}, \overline{\bf 3}, \overline{\bf 3}, \overline{\bf 3}, {\bf 1}\}$,
respectively.
The symmetric representation ${\bf 6}$,
for example, is forbidden by antisymmetry. The states in Eq.~(\ref{eq:states}) map naturally to the eight states of a single qu8it.
Further details about this embedding can be found in App.~\ref{ap:embedQaQ}.

The anti-quarks are mapped to qu8its in an analogous way to the quarks, but with the replacement 
$\{r,g,b\} \rightarrow \{\overline{r}, \overline{g}, \overline{b}\}$,
\begin{eqnarray}
    \big\{  | \overline{\rm qu8it}\rangle  \big\}
    \ & = & \
    \big\{ 
    |\Omega\rangle \ , \ 
    |\overline{q}_r\rangle\ , \ 
    |\overline{q}_g\rangle\ , \ 
    |\overline{q}_b\rangle\ , \ 
    |\overline{q}_g \overline{q}_b\rangle\ , \ 
    -|\overline{q}_r \overline{q}_b\rangle\ , \ 
|\overline{q}_r \overline{q}_g\rangle\ , \ 
    |\overline{q}_r \overline{q}_g \overline{q}_b\rangle
    \big\}
        \nonumber\\
    \ & = & \
    \big\{ 
    |\Omega\rangle \ , \ 
    \hat{c}_{\overline{r}}^\dagger|\Omega\rangle\ , \ 
    \hat{c}_{\overline{g}}^\dagger|\Omega\rangle\ , \ 
    \hat{c}_{\overline{b}}^\dagger|\Omega\rangle\ , \ 
    \hat{c}_{\overline{g}}^\dagger \hat{c}_{\overline{b}}^\dagger|\Omega\rangle\ , \ 
    -\hat{c}_{\overline{r}}^\dagger \hat{c}_{\overline{b}}^\dagger|\Omega\rangle\ , \ 
    \hat{c}_{\overline{r}}^\dagger \hat{c}_{\overline{g}}^\dagger|\Omega\rangle\ , \ 
    \hat{c}_{\overline{r}}^\dagger \hat{c}_{\overline{g}}^\dagger \hat{c}_{\overline{b}}^\dagger|\Omega\rangle  
    \big\}
    \nonumber\\
    \ & = & \ 
    \big\{ 
    |\bar 1\rangle \ , \ 
    |\bar 2\rangle\ , \ 
    |\bar 3\rangle\ , \ 
    |\bar 4\rangle\ , \ 
    |\bar 5\rangle\ , \ 
    |\bar 6\rangle\ , \ 
    |\bar 7\rangle\ , \ 
    |\bar 8\rangle
    \big\} 
    \ .
\label{eq:states2}
\end{eqnarray}
As is the case for the quark mapping, each state in Eq.~\eqref{eq:states2} transforms as a single SU(3) irrep,
which for the anti-quarks are
$\{{\bf 1}, \overline{\bf 3}, \overline{\bf 3}, \overline{\bf 3}, {\bf 3}, {\bf 3}, {\bf 3}, {\bf 1}\}$, respectively.

\begin{figure}[!h]
    \centering
    \includegraphics[width=0.65\columnwidth]{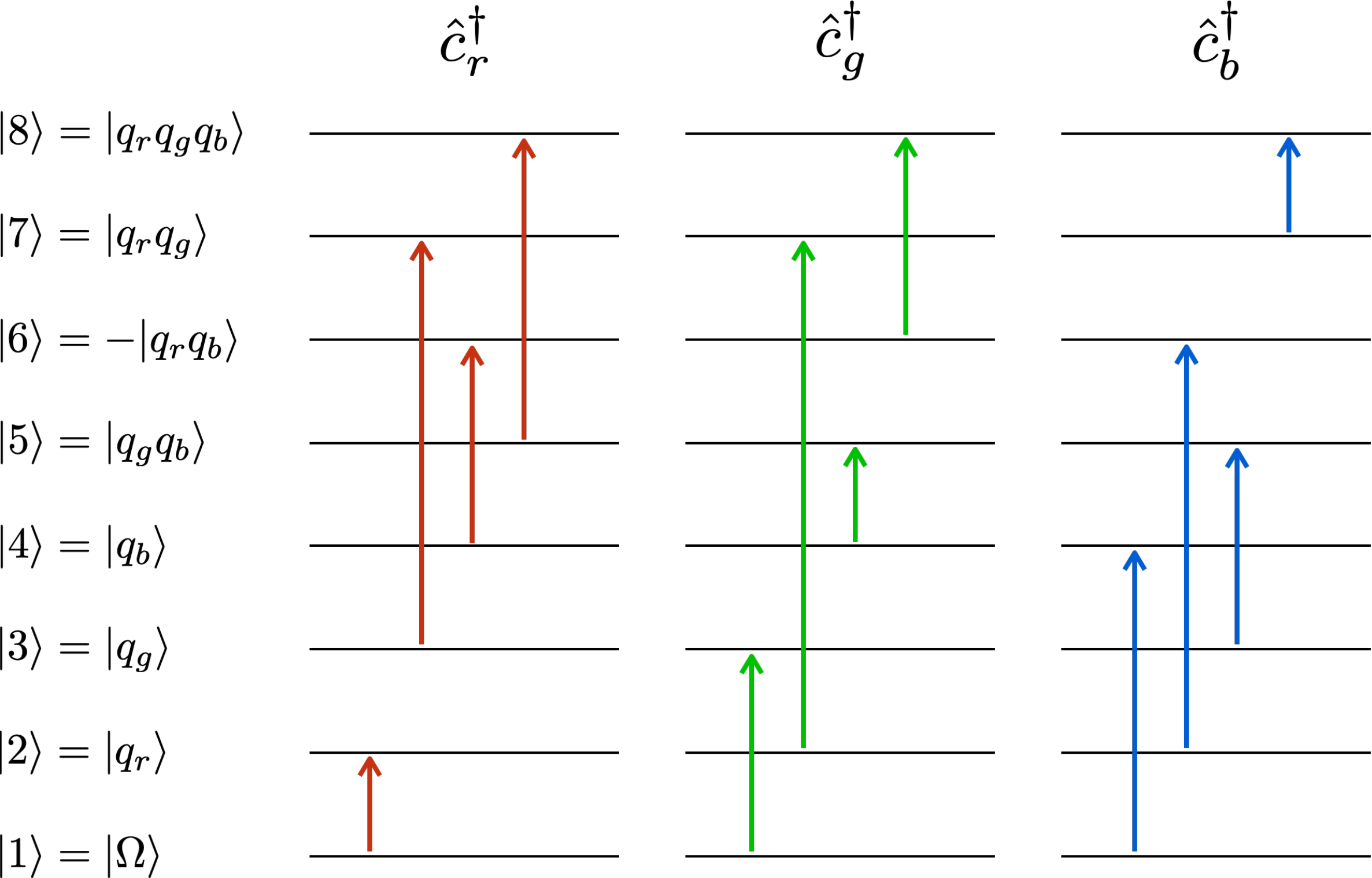}
    \caption{Transitions among the quark states mapped to 
    a qu8it, defined in Eq.~\eqref{eq:states}, 
    induced by the creation operators $\hat c_r^\dagger$ (left),  $\hat c_g^\dagger$ 
    (center) and $\hat c_b^\dagger$ (right) from Eq.~(\ref{eq:cmats}).
    The color of the arrows corresponds to the color charge of the operator.
    } 
    \label{fig:rgbTrans}
\end{figure}

With the mappings defined in Eqs.~\eqref{eq:states} and \eqref{eq:states2}, 
the relevant operators for quantum simulation can be formed.
The color-charge operator acting on the quarks is
\begin{eqnarray}
\hat Q^{(a)} \ & = & \ \hat {\bf c}^\dagger T^a \hat {\bf c}
\ ,\ \ 
\hat {\bf c}\ =\ \left( \hat c_r, \hat c_g, \hat c_b \right)^T
    \ ,
    \label{eq:chargeTa}
\end{eqnarray}
As the $\overline{\bf 3}$ of di-quarks is also present in the mapping to qu8its, 
the color-charge operators acting on the 
anti-fundamental representation (the same as the anti-quarks) is also required,
\begin{eqnarray}
\hat {\bar Q}^{(a)} \ & = & \ {\hat {\bar{\bf c}}}^\dagger \bar T^a {\hat {\bar {\bf c}}}
\ ,\ \ 
{\hat {\bar {\bf c}}}\ = \ \left( \hat c_{\overline{r}} , \hat c_{\overline{g}}, \hat c_{\overline{b}} \right)^T
\ ,\ \ 
\bar T^a \ = \ \left(-T^a\right)^*
    \ .
\label{eq:antichargeTa}
\end{eqnarray}
Therefore, the ($8\times 8$) matrix representations of the 
color-charge operator acting on qu8its and anti-qu8its are
\begin{eqnarray}
\hat Q^{(a)}  & \ \rightarrow  \ &
\widetilde{Q}^{(a)}\ =\ 
\left(
{\setstretch{1.3}
    \begin{array}{@{}c|c|c|c@{}}
    0 &0 &0 &0 \\    \hline 
0 &T^a &0 &0  \\  \hline 
0 &0 &\bar T^a  & 0  \\  \hline 
0 &0 &0 &0 
    \end{array}
    }
    \right)
    \ , \ \   
\hat{\bar{Q}}^{(a)}  \ \rightarrow  \ 
\widetilde{\bar{Q}}{}^{(a)}\ =\ 
\left(
{\setstretch{1.3}
    \begin{array}{@{}c|c|c|c@{}}
    0 &0 &0 &0 \\    \hline 
0 &\bar{T}^a &0 &0  \\  \hline 
0 &0 & T^a  & 0  \\  \hline 
0 &0 &0 &0 
    \end{array}
    }
    \right)
    \ ,
\label{eq:chargeMat}
\end{eqnarray}
respectively, where the blocks of $\widetilde{Q}^{(a)}$ in Eq.~\eqref{eq:chargeMat} correspond to the action on
$\{{\bf 1}, {\bf 3}, \overline{\bf 3}, {\bf 1}\}$ irreps, 
and the blocks of $\widetilde{\bar Q}{}^{(a)}$ correspond to the action on 
$\{{\bf 1}, \overline{\bf 3}, {\bf 3}, {\bf 1} \}$ irreps.
A qu8it prepared in an arbitrary state $|\psi\rangle$, acted on by the  
color-charge operator, becomes
\begin{eqnarray}
|\psi\rangle \ & = & \
 \big\{  |{\rm qu8it}\rangle  \big\}\cdot {\bm \xi}\ =\ 
\sum_{k=1}^8 \ |k\rangle \ \xi_k
\ , \ \ 
\hat Q^{(a)} |\psi\rangle \ =\  
 \big\{  |{\rm qu8it}\rangle  \big\}\cdot \widetilde{Q}^{(a)} \cdot {\bm \xi}\ =\ 
\sum_{k=1}^8  \ |k\rangle\ 
\left(\widetilde{Q}^{(a)} \cdot {\bm \xi} \right)_k
    \ ,
\label{eq:q8i}
\end{eqnarray}
where ${\bm \xi}$ is the vector of complex numbers defining the state.
The baryon-number operator has a diagonal matrix representation,
\begin{eqnarray}
\hat{\mathcal{B}} \rightarrow \widetilde{\mathcal{B}}
\ & =  & \
\frac{1}{3} {\rm diag}
\left( 
0,1,1,1,2,2,2,3
\right)
    \ .
\label{eq:Bop}
\end{eqnarray}
Finally, the
annihilation and creation operators 
acting on a qu8it have matrix representations
\begin{eqnarray}
\widetilde{c_r} & \ = \ & 
\left(
{\setstretch{1}
    \begin{array}{@{}cccccccc@{}}
    0 & 1 &0 &0 &0 &0 &0 &0 \\   
    0 & 0 &0 &0 &0 &0 &0 &0 \\   
    0 & 0 &0 &0 &0 &0 &1 &0 \\   
    0 & 0 &0 &0 &0 &-1 &0 &0 \\   
    0 & 0 &0 &0 &0 &0 &0 &1 \\   
    0 & 0 &0 &0 &0 &0 &0 &0 \\   
    0 & 0 &0 &0 &0 &0 &0 &0 \\   
    0 & 0 &0 &0 &0 &0 &0 &0 \\   
    \end{array}
    }
    \right)
    \ ,\ \ 
\widetilde{c_g} \ =\  
\left(
{\setstretch{1}
    \begin{array}{@{}cccccccc@{}}
    0 & 0 &1 &0 &0 &0 &0 &0 \\   
    0 & 0 &0 &0 &0 &0 &-1 &0 \\   
    0 & 0 &0 &0 &0 &0 &0 &0 \\   
    0 & 0 &0 &0 &1 &0 &0 &0 \\   
    0 & 0 &0 &0 &0 &0 &0 &0 \\   
    0 & 0 &0 &0 &0 &0 &0 &1 \\   
    0 & 0 &0 &0 &0 &0 &0 &0 \\   
    0 & 0 &0 &0 &0 &0 &0 &0 \\   
    \end{array}
    }
    \right)
    \ ,\ \ 
    \widetilde{c_b} \ =\  
\left(
{\setstretch{1}
    \begin{array}{@{}cccccccc@{}}
    0 & 0 &0 &1 &0 &0 &0 &0 \\   
    0 & 0 &0 &0 &0 &1 &0 &0 \\   
    0 & 0 &0 &0 &-1 &0 &0 &0 \\   
    0 & 0 &0 &0 &0 &0 &0 &0 \\   
    0 & 0 &0 &0 &0 &0 &0 &0 \\   
    0 & 0 &0 &0 &0 &0 &0 &0 \\   
    0 & 0 &0 &0 &0 &0 &0 &1 \\   
    0 & 0 &0 &0 &0 &0 &0 &0 \\   
    \end{array}
    }
    \right)
   \ ,
\label{eq:cmats}
\end{eqnarray}
respectively, with 
$\widetilde{c_r}^\dagger = \widetilde{c_r}^T$, 
$\widetilde{c_g}^\dagger =  \widetilde{c_g}^T$, and 
$\widetilde{c_b}^\dagger = \widetilde{c_b}^T$,
satisfying 
the (required) fermionic anti-commutation relations, $\{\widetilde{c_\alpha},\widetilde{c_\beta\,}^\dag\}=\delta_{\alpha\beta}$ and $\{\widetilde{c_\alpha},\widetilde{c_\beta\,}\}=\{\widetilde{c_\alpha}^\dag,\widetilde{c_\beta\,}^\dag\}=0$ (with $\alpha,\, \beta \in \{ r,g,b \}$).
The annihilation and creation operators act on the states analogously to the charge operators.
For example, the action of $\hat{c}_r$ on a state $|\psi\rangle$ is
\begin{eqnarray}
|\psi\rangle \ & = & \
 \big\{  |{\rm qu8it}\rangle  \big\}\cdot {\bm \xi}
\ ,\ \ 
\hat c_r |\psi\rangle \ =\  
 \big\{  |{\rm qu8it}\rangle  \big\}\cdot \widetilde{c_r} \cdot {\bm \xi}   
 \ .
\label{eq:cr8i}
\end{eqnarray}
The actions of the creation operators 
in Eq.~(\ref{eq:cmats})
are shown in Fig.~\ref{fig:rgbTrans}.
The creation and annihilation operators acting on the anti-qu8it have the same representations as those acting on the qu8it in Eq.~\eqref{eq:cmats}: $\widetilde{\bar{c}_\alpha} = \widetilde{c_\alpha}$. 
%

\section{The 1+1D QCD Hamiltonian Mapped to Qu8its}
\label{sec:Hami}
\begin{figure}[h!]
    \centering
    \includegraphics[width=0.75\textwidth]{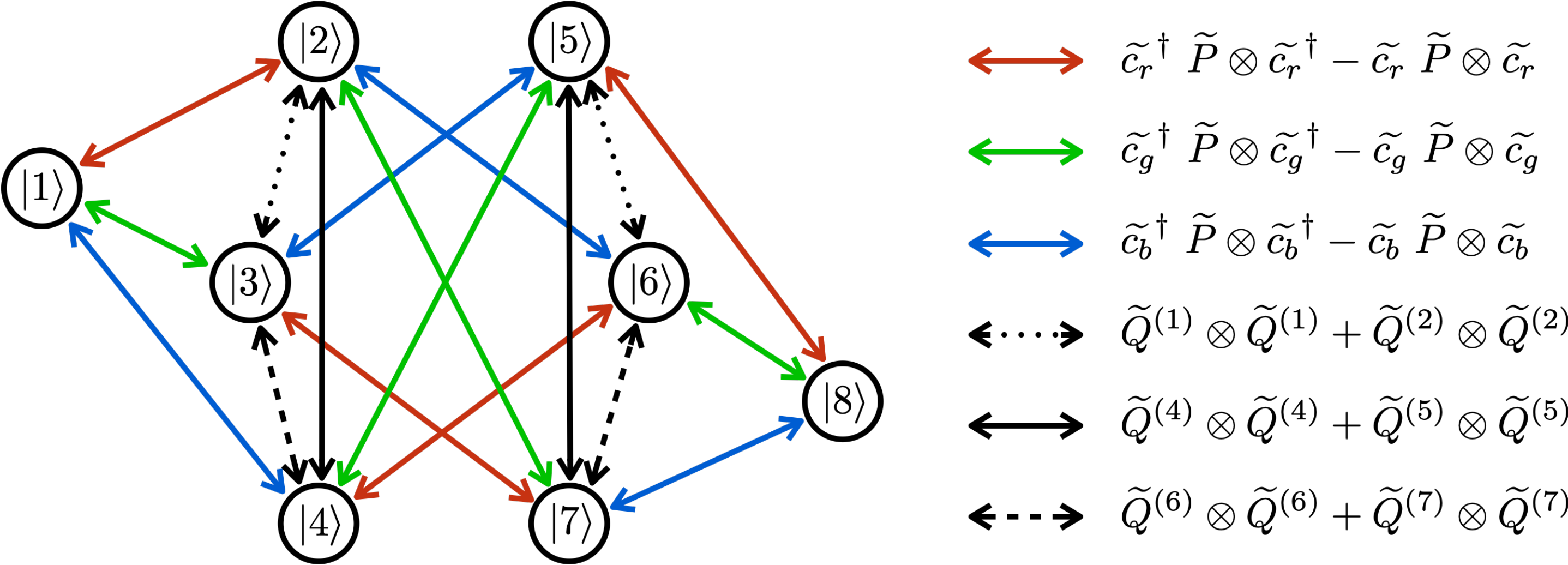}
    \caption{A connectivity map among the eight qu8it states that is required by the 
    Hamiltonian in Eq.~(\ref{eq:Hq8mat}). 
    Colored connections are for the kinetic term, while black ones are for the color charge-charge interactions (different line styles correspond to different charge combinations).}
    \label{fig:ConnectStates}
\end{figure}

\noindent
In terms of the annihilation and creation operators, the 
$N_f=1$ SU(3) Hamiltonian 
in Eq.~(\ref{eq:GFHam}) can be written as 
\begin{eqnarray}
\hat H \ & = & \ \hat H_{kin} \ +\ \hat H_{m}\ +\ \hat H_{el}
\nonumber\\
\ & = & \
\frac{1}{2}\  
\sum_{n=0}^{2L-2}\ \sum_{\alpha=r,g,b}
\left( 
\hat c_{\alpha,n}^\dagger \hat c_{\alpha,n+1}^\dagger - \hat c_{\alpha,n} \hat c_{\alpha,n+1}
\right)
\ +\ 
3 m\  \sum_{n=0}^{2L-1}\ \hat{\mathcal{B}}_n
\ +\ 
\frac{g^2}{2}\  
\sum_{n=0}^{2L-2}\ 
\sum_{a=1}^8\ 
\left( \sum_{m\le n} \hat Q^{(a)}_m \right)^2
\ .
\label{eq:Hq8}
\end{eqnarray}
When written explicitly in terms of matrix-operators acting on qu8its
(analogous to the JW mapping to qubits with the operators written in terms of Pauli matrices),
it becomes
\begin{eqnarray}
\hat H \ & \rightarrow & \
\frac{1}{2}\  
\sum_{n=0}^{2L-2}\ \sum_{\alpha=r,g,b}
\left[
\left(\widetilde{c_\alpha}^\dagger\ \widetilde{P}\right)_n \otimes \ {\widetilde{c_\alpha}_{, n+1}}^\dagger
\ -\ 
\left(\widetilde{c_\alpha}\ \widetilde{P}\right)_n \otimes\  \widetilde{c_\alpha}_{, n+1}
\right]
\ +\ 
3 m\  \sum_{n=0}^{2L-1}\ \widetilde{\mathcal{B}}_n \nonumber \\
& & + \ 
\frac{g^2}{2}\  
\sum_{n=0}^{2L-2}\ 
\sum_{a=1}^8\ 
\left( 
\sum_{m\in{\rm even}}^n \widetilde{Q}^{(a)}_m 
\ + \sum_{m\in{\rm odd}}^n \widetilde{\bar Q}{}^{(a)}_m
\right)^2
    \ ,
\label{eq:Hq8mat}
\end{eqnarray}
where the phase matrix $\widetilde{P}$ is 
\begin{eqnarray}
\hat P \ \rightarrow \ \widetilde{P} \ & = & \ {\rm diag}\left(1,-1,-1,-1,1,1,1,-1\right)
    \ .
\label{eq:phase}
\end{eqnarray}
The $\widetilde{P}$ matrix is the generalized form of the $\sigma^z$ Pauli matrix,
generally acting on strings of qudits between quark operators in order to satisfy Fermi statistics.
Examining the connectivity map shown in Fig.~\ref{fig:ConnectStates},
one observes that each state is connected to either 3 or 5 other states within the qu8it.
The states corresponding to color-singlet states have connectivity of 3 
(originating from the kinetic term), 
while those corresponding to color-triplet or anti-triplet states have connectivity of 5 
(originating from the kinetic and chromo-electric terms).

The extension to systems 
with a
larger number of quark flavors,
$N_f>1$, is straightforward.
For each flavor of quark at a given lattice site, there is a corresponding qu8it, 
and similarly for anti-quark sites.
The kinetic and mass terms in the Hamiltonian are replicated for each flavor (with the appropriate masses), and the color-charge operators are extended as in Eq.~(\ref{eq:SU3charges}).

\subsection{QCD with $N_f=1$ on $L=1$ Spatial Site with OBCs}
\label{sec:one}
\noindent
It is helpful to explore examples of 
mappings to qu8its.
Consider $L=1$, with lattice sites $n=0,1$, 
with one flavor $N_f=1$, 
which maps to two qu8its (one for the quarks and one for the anti-quarks),
a system that we have studied previously~\cite{Farrell:2022wyt}.
The matrix representation of the Hamiltonian, as given in Eq.~(\ref{eq:Hq8mat}),
reduces to
\begin{eqnarray}
H_1 \ & = & \
\frac{1}{2}\left(
\widetilde{c_{r}}^\dagger\ \widetilde{P} \otimes \widetilde{c_{r}}^\dagger
+\widetilde{c_{g}}^\dagger\ \widetilde{P} \otimes \widetilde{c_{g}}^\dagger
+\widetilde{c_{b}}^\dagger\ \widetilde{P} \otimes \widetilde{c_{b}}^\dagger
-\widetilde{c_{r}}\ \widetilde{P} \otimes \widetilde{c_{r}}
-\widetilde{c_{g}}\ \widetilde{P} \otimes \widetilde{c_{g}}
-\widetilde{c_{b}}\ \widetilde{P} \otimes \widetilde{c_{b}}
\right)
\nonumber\\
&  & + \
3 m\  
\left(
\widetilde{\mathcal{B}} \otimes \widetilde{I} 
+ \widetilde{I} \otimes \widetilde{\mathcal{B}}
\right)
\ +\ 
\frac{g^2}{2}\ \sum_a\ \left( \widetilde{Q}^{(a)}\otimes \widetilde{I} \right)^2
\ +\ 
\frac{h^2}{2}\ \sum_a\ \left( 
\widetilde{Q}^{(a)}\otimes \widetilde{I}
+ \widetilde{I} \otimes\widetilde{\bar Q}^{(a)}  \right)^2 \ ,
\nonumber\\
\ & = & \
H_{1kin}\ +\ H_{1m}\ +\ H_{1el} \ +\ H_{1h}
    \ ,
\label{eq:H1}
\end{eqnarray}
where $\widetilde{I}$ is the identity operator.
The term with coefficient $h$ has been included to enforce color-neutrality across the lattice as $h\rightarrow\infty$, 
as we implemented in previous work~\cite{Farrell:2022wyt}.
This generates a significant penalty for chromo-electric-energy density beyond the end of the spatial 
lattice, and without this term color-edge states appear as low-lying states in the spectrum due to OBCs~\cite{Farrell:2022wyt}.
In the large-$h$ limit, only color-singlet states remain at low energies.

This system is sufficiently simple and of small dimensionality,
involving a $64\times 64$ Hamiltonian matrix,
that it can be exactly diagonalized with classical computers.
Projecting to states with good baryon number further reduces the size of the matrix.
For example, 
in the $B=0$ sector, the contributing configurations correspond to
i) both qu8its in the vacuum (a ${\bf 1}$);
ii) the qu8its are in the one-quark one-anti-quark sector
(${\bf 3}\otimes\overline{\bf 3}={\bf 8}\oplus {\bf 1}$);
iii) the qu8its are in the two-quark two-anti-quark sector, 
($\overline{\bf 3}\otimes {\bf 3}={\bf 8}\oplus {\bf 1}$); and,
iv) both qu8its are in the completely occupied state, a baryon-anti-baryon pair (a ${\bf 1}$).
Consequently, the total number of $B=0$ basis states is $n_{B=0}=1+9+9+1 = 20$.
However, a large value of $h$ propels the ${\bf 8}$'s high in the spectrum, leaving only four color-singlet states in the low-lying spectrum.  
These are formed from linear combinations of the eight pairings of states in the qu8its.\footnote{If we were working in U(1) lattice gauge theory describing quantum electrodynamics, 
the situation would be somewhat less complex because each (tensor-product) basis state is an eigenstate of the electric-charge operator.  
This is not the situation for non-Abelian theories, where the color-charge operator generally mixes basis states.}

As this is a system we have analyzed previously using the JW mapping to qubits~\cite{Farrell:2022wyt}, 
the low-lying spectra and time-evolution from arbitrary initial states are known.
The (exact) time evolution found from matrix exponentiation of the Hamiltonian in Eq.~\eqref{eq:H1},
is found to furnish results that agree with our previous analyses.

As shown in Eq.~\eqref{eq:Hg21} below, the chromo-electric term $\hat{H}_{1el}$ is diagonal in the qu8it computational basis for $L=1$. 
Thus, in the case of an ideal quantum computer, with an initial state that is a color-singlet, exact time evolution will leave the system in a color-singlet state at all subsequent times, 
even without the ``$h$-term'' in Eq.~\eqref{eq:H1}.
As such, that term can be omitted in the time-evolution operator in the case of $L=1$.  
For systems with $L>1$, however,
color-charge is violated by Trotterized time evolution 
(in particular, due to Trotterization of the eight contributions to the color sum 
in the chromo-electric field term, when the color-charge operators act on different sites),
and, consequently, including the ``$h$-term'' is a means to mitigate this violation.

\subsubsection{Quantum Circuits and Givens Rotations}
\label{sec:givens}
\noindent
To establish the quantum circuits for the $N_f=1$ and $L=1$ system, 
the Hamiltonian is decomposed into unitary operations on each of the qu8its.
As is standard for quantum operations on qudits, 
the Hamiltonian in Eq.~(\ref{eq:H1}) is decomposed into generators of Givens rotations,
${\cal X}_{ij}$,  ${\cal Y}_{ij}$, and ${\cal Z}_{i}$.
In this case, $d=8$, 
these provide a complete set of generators for 
SU(8) transformations on a single qu8it.
It is convenient, in an effort to better connect with hardware aspects of simulations, to 
work with Hadamard-Walsh matrices, $w_i$,  rather than with ${\cal Z}_{i}$.
For $d=8$, there are 28 ${\cal X}_{ij}$, 28 ${\cal Y}_{ij}$, and 8 $w_{i}$ (including the identity), 
as defined in
App.~\ref{ap:su8givens}.

The kinetic-energy operator in the Hamiltonian in Eq.~\eqref{eq:H1} 
is comprised only of terms that act on both qu8its.
As the annihilation and creation operators induce multiple transitions within a qu8it, as depicted in Fig.~\ref{fig:rgbTrans},
this term decomposes into multiple tensor products of Givens matrices,
\begin{eqnarray}
H_{1kin} \ & = & \
\frac{1}{4} 
\sum_{\substack{ r \in \{(12),(13),(14), \\ 
\ \ (58),(68),(78), \\
\ \ -(26),-(27),-(35), \\ \ \ 
-(37), -(45), -(46)
\}}}
\left( {\cal X}_r\otimes {\cal X}_r -  {\cal Y}_r\otimes {\cal Y}_r  \right)
\nonumber\\
&  &  + \
\frac{1}{4} 
\sum_{\substack{ (r,s) \in  \{
(12)(58), (13)(68), \\
\ \ (14)(78), (26)(35), \\
\ \ (27)(45), (37)(46)
\}}}
\left( {\cal X}_r\otimes {\cal X}_s + {\cal X}_s\otimes {\cal X}_r 
-  {\cal Y}_r\otimes {\cal Y}_s -  {\cal Y}_s\otimes {\cal Y}_r  \right)
\nonumber\\
& & + \
\frac{1}{4} 
\sum_{\substack{ (r,s) \in  \{
(12)(37), (46)(12), \\
\ \ (27)(13), (13)(45), (14)(26), \\
\ \ (35)(14), (78)(26), (27)(68), \\
\ \ (35)(78), (58)(37), (68)(45), \\
\ \ (46)(58)\}}}
\left( {\cal X}_r\otimes {\cal X}_s - {\cal X}_s\otimes {\cal X}_r 
-  {\cal Y}_r\otimes {\cal Y}_s +  {\cal Y}_s\otimes {\cal Y}_r  \right)
\ ,
\label{eq:H1K}
\end{eqnarray}
where the minus sign in some of the indices in the first summation means a global minus sign for that corresponding term.
The mass term in the Hamiltonian is diagonal, given in terms of the action of the baryon-number matrix $\widetilde{\mathcal{B}}$ defined in Eq.~(\ref{eq:Bop}),
and which can be further decomposed into the $w_i$,
\begin{eqnarray}
H_{1m} \ & = & \
3 m\  
\left(
\widetilde{\mathcal{B}} \otimes \widetilde{I} 
+ \widetilde{I} \otimes \widetilde{\mathcal{B}}
\right)
\nonumber\\
& = & \
\frac{m}{\sqrt{2}}
\left(
(6 w_1 - 3 w_2-w_4-w_6-w_8)\otimes \widetilde{I} 
\ +\ 
\widetilde{I} \otimes(6 w_1 - 3 w_2-w_4-w_6-w_8)
\right)
\ ,
\nonumber\\
\widetilde{\mathcal{B}} \ & = & \ \frac{1}{3} {\rm diag} \left( 0,1,1,1,2,2,2,3\right)
\ =\ 
\frac{1}{3\sqrt{2}} (6 w_1 - 3 w_2-w_4-w_6-w_8)
\  .
\label{eq:Hmass1}
\end{eqnarray}
For a single site, the contribution of the chromo-electric-energy density to the Hamiltonian 
from within the lattice,
receiving contributions from the $n=0$ qu8it site only,
can be written as (details can be found in App.~\ref{ap:QQterms}),
\begin{eqnarray}
H_{1el} & = & 
\frac{g^2}{2}\ \sum_a\ \left( \widetilde{Q}^{(a)}\otimes \widetilde{I} \right)^2
\ \rightarrow\ 
\frac{2g^2}{3}
{\rm diag}\left( 0,1,1,1,1,1,1,0 \right) \otimes \widetilde{I} 
\nonumber\\
& = & 
\frac{g^2}{2}\ \left(
8 w_1\otimes w_1 - \frac{8}{3} \left( w_3 + w_5 + w_7 \right) \otimes w_1
 \right)
\nonumber\\
& = & 
\frac{g^2}{2}\ \left( 
\widetilde{I} \otimes \widetilde{I}  
- {\rm diag}\left(1,-\frac{1}{3},-\frac{1}{3},-\frac{1}{3},-\frac{1}{3},-\frac{1}{3},-\frac{1}{3},1\right)\otimes \widetilde{I} 
\right)
\label{eq:Hg21}
\ \ .
\end{eqnarray}
The $h$-term introduced to enforce color neutrality of 
an $L>1$ lattice, requires 
computing
the square of the total color charge of the lattice (see App.~\ref{ap:QQterms}),
\begin{eqnarray}
H_{1h} & = & 
\frac{h^2}{2}\ \sum_{a=1}^8\ \left( 
\widetilde{Q}^{(a)}\widetilde{Q}^{(a)}\otimes \widetilde{I}
+ \widetilde{I} \otimes \widetilde{\bar Q}{}^{(a)}\widetilde{\bar Q}{}^{(a)}  
+ 2 \widetilde{Q}^{(a)}\otimes\widetilde{\bar Q}{}^{(a)}
\right)
\nonumber\\
& = & 
\frac{h^2}{2}\ 
\left( 16 w_1\otimes w_1 
- \frac{8}{3} w_1\otimes (w_3+w_5+w_7)
- \frac{8}{3} (w_3+w_5+w_7)\otimes w_1
\right.
\nonumber\\
& & 
\left. 
- \left(w_3-w_5\right)\otimes \left(w_3-w_5\right)
- \frac{1}{3} 
\left(w_4+w_6-2 w_8\right) \otimes \left(w_4+w_6-2 w_8\right)
\right.
\nonumber\\
& & 
\left. 
+
\frac{1}{2} 
\sum_{\substack{ r \in \{(23),(24),(34), \\ 
\ \ (56),(57),(67)\}}}
\left(  {\cal Y}_r\otimes {\cal Y}_r - {\cal X}_r\otimes {\cal X}_r \right)
\right.
\nonumber\\
& & 
\left. 
+
\frac{1}{2} 
\sum_{\substack{ (r,s) \in  \{(23)(56), \\ \ \ \ \ 
(24)(57), (34)(67)\}}}
\left(
{\cal X}_r\otimes {\cal X}_s + {\cal Y}_r\otimes {\cal Y}_s
+
{\cal X}_s\otimes {\cal X}_r + {\cal Y}_s\otimes {\cal Y}_r
\right)
\right)
\ \ .
\label{eq:H1h}
\end{eqnarray}
\begin{figure}[th!]
    \centering
    \includegraphics[width=0.43\textwidth]{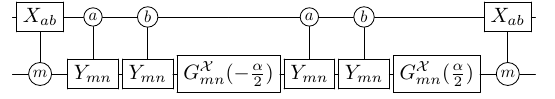}\ \ \ \ 
    \includegraphics[width=0.43\textwidth]{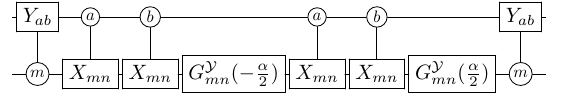}
    \caption{Quantum circuits acting on qu8its that implement (left panel)
    $G^{{\cal X}{\cal X}}_{abmn}(\alpha) = e^{-i\alpha {\cal X}_{ab}\otimes {\cal X}_{mn}}$,
    and (right panel) 
     $G^{{\cal Y}{\cal Y}}_{abmn}(\alpha) = e^{-i\alpha {\cal Y}_{ab}\otimes {\cal Y}_{mn}}$.
     These are required to implement Trotterized time evolution, along with single-qu8it gates,     
     corresponding to the Hamiltonian terms in Eqs.~(\ref{eq:H1K}), (\ref{eq:Hmass1}), (\ref{eq:Hg21}) 
     and (\ref{eq:H1h}).   
     This circuit structure is reproduced from our previous work~\cite{Illa:2023scc}.
     }
    \label{fig:EntanglingGateCircuit}
\end{figure}
The Trotterized time evolution arising from this Hamiltonian, 
written in terms of generators of Givens transformation, 
can be implemented on qu8its using known quantum circuits,
requiring 96 two-qu8it Givens rotations, 
and including the  $H_{1h}$ contribution would require an additional 26 
rotations.\footnote{We had hoped that the qu8it mapping would eliminate or mitigate the violation 
of SU(3) color charge occurring 
for $L>1$ lattices
due to Trotterization (that we identified in Ref.~\cite{Farrell:2022wyt}).
However, this is not the case, and Trotterization 
of the time-evolution arising from the qu8its Hamiltonian also violates color-charge conservation for $L>1$.}
Figure~\ref{fig:EntanglingGateCircuit} shows the quantum circuits for implementing the 
unitary operations of the form
$G^{{\cal X}{\cal X}}_{abmn}(\alpha) = e^{-i\alpha {\cal X}_{ab}\otimes {\cal X}_{mn}}$ and 
$G^{{\cal Y}{\cal Y}}_{abmn}(\alpha) = e^{-i\alpha {\cal Y}_{ab}\otimes {\cal Y}_{mn}}$,
which are the only required entangling structures, as a function of controlled gates.
The associated gate-count for implementing one Trotter step of time evolution for $N_f=1$ and $L=1$
is, including the $h$-term, 732 controlled-$X$ and -$Y$ gates and 249 single qu8it rotations, which is much larger compared to the direct application of $G^{{\cal X}{\cal X}}_{abmn}(\alpha)$ and $G^{{\cal Y}{\cal Y}}_{abmn}(\alpha)$ via the generalized M\o{}lmer-S\o{}rensen (MS) gates~\cite{Molmer:1998mq,Low:2019}.

\subsubsection{Restructuring}
\label{sec:restruct}
\noindent
A recent paper~\cite{Calajo:2024qrc}, 
where a different route to implementing similar types of gates is taken, 
indicates significant reductions in two-qu8it rotations can be gained by grouping operators.
This is motivated by advances in quantum hardware~\cite{Low:2019,Ringbauer:2021lhi} so that multiple of the transitions associated with, for instance, the addition of a red-quark at a given site can be implemented simultaneously, as opposed to sequentially.  
This suggests that all of the operations associated with the red-quark-creation operator 
can be grouped together.  
Similarly for the green and blue operations, and for their corresponding annihilation.
The kinetic-energy term in Eq.~(\ref{eq:H1K}) can be greatly simplified and 
written as
\begin{align}
H_{1kin} &= \frac{1}{4}(A_0^{(r)}\otimes A_1^{(r)}-B_0^{(r)}\otimes B_1^{(r)})+\frac{1}{4}(A_0^{(g)}\otimes A_1^{(g)}-B_0^{(g)}\otimes B_1^{(g)})+\frac{1}{4}(A_0^{(b)}\otimes A_1^{(b)}-B_0^{(b)}\otimes B_1^{(b)})\ ,
\label{eq:hoppregroup}
\end{align}
where
\begin{eqnarray}
    A_0^{(r)} & = & {\cal X}_{(12)}-{\cal X}_{(37)}+{\cal X}_{(46)}+{\cal X}_{(58)} 
    \ , \ \ A_1^{(r)}={\cal X}_{(12)}+{\cal X}_{(37)}-{\cal X}_{(46)}+{\cal X}_{(58)}\ , 
    \nonumber \\
    A_0^{(g)}& = & {\cal X}_{(13)}+{\cal X}_{(27)}-{\cal X}_{(45)}+{\cal X}_{(68)} 
    \ , \ \ A_1^{(g)}={\cal X}_{(13)}-{\cal X}_{(27)}+{\cal X}_{(45)}+{\cal X}_{(68)}\ , 
    \nonumber \\
    A_0^{(b)}& = & {\cal X}_{(14)}-{\cal X}_{(26)}+{\cal X}_{(35)}+{\cal X}_{(78)} 
    \ , \ \ A_1^{(b)}={\cal X}_{(14)}+{\cal X}_{(26)}-{\cal X}_{(35)}+{\cal X}_{(78)}\ ,
    \label{eq:A12def}
\end{eqnarray}
and the $B^{(\alpha)}_n$ operators are analogous to the  
$A^{(\alpha)}_n$ operators with ${\cal X}\leftrightarrow {\cal Y}$.
The 96 two-qu8it entangling terms in Eq.~(\ref{eq:H1K}) are reduced to just 6 in Eq.~(\ref{eq:hoppregroup}).
The mass term is unchanged, as is the contribution from the chromo-electric field proportional to $g^2$,
neither of which involve two-qu8it entangling gates.

For the $h$-term, a similar simplification of terms exists by groupings into commuting sets,
\begin{eqnarray}
H_{1h} & = & 
\frac{h^2}{2}\  \left(
    \frac{1}{2}( D^{(12)} \otimes D^{(12)} - C^{(12)} \otimes C^{(12)})
    +\frac{1}{2}( D^{(45)} \otimes D^{(45)} - C^{(45)} \otimes C^{(45)}) 
    \right.
    \nonumber \\ 
    & & + \ \frac{1}{2}(  D^{(67)} \otimes D^{(67)} - C^{(67)} \otimes C^{(67)})
     + 2 \widetilde{Q}^{(3)}\otimes \widetilde{\bar Q}{}^{(3)} 
     +2 \widetilde{Q}^{(8)}\otimes \widetilde{\bar Q}{}^{(8)}
    \nonumber \\ 
    & & +  \left.
2 \widetilde{I} \otimes \widetilde{I}  
- {\rm diag}\left(1,-\frac{1}{3},-\frac{1}{3},-\frac{1}{3},-\frac{1}{3},-\frac{1}{3},-\frac{1}{3},1\right)\otimes \widetilde{I} 
- \widetilde{I} \otimes  {\rm diag}\left(1,-\frac{1}{3},-\frac{1}{3},-\frac{1}{3},-\frac{1}{3},-\frac{1}{3},-\frac{1}{3},1\right)
\right)
    \ .
    \label{eq:elregroup}
\end{eqnarray}
where the $C$ and $D$ terms are, respectively,
\begin{eqnarray}
    C^{(12)} & = &  {\cal X}_{(23)}-{\cal X}_{(56)} \ , \ \
    C^{(45)} \ =\  {\cal X}_{(24)}-{\cal X}_{(57)} \ , \ \
    C^{(67)} \ =\  {\cal X}_{(34)}-{\cal X}_{(67)} \ , 
    \nonumber\\
     D^{(12)} & = & {\cal Y}_{(23)}+{\cal Y}_{(56)} \ , \ \
     D^{(45)} \ = \ {\cal Y}_{(24)}+{\cal Y}_{(57)} \ , \ \
     D^{(67)} \ = \ {\cal Y}_{(34)}+{\cal Y}_{(67)} 
    \ .
\label{eq:ABH1hred}
\end{eqnarray}
The 26 two-qu8it operations in $H_{1h}$ 
in Eq.~(\ref{eq:H1h}) is reduced to 8 in Eq.~(\ref{eq:elregroup})
by this restructuring.

\subsection{QCD with $N_f$ Flavors on $L$ Spatial Site and Quantum Resource Requirements}
\label{sec:two}
\noindent
Considering the  general situation of 
an arbitrary number of spatial lattice sites 
and an arbitrary number of flavors of quarks, the Hamiltonian becomes (including the $h$-term),
\begin{eqnarray}
\hat H \ & = & \ \hat H_{kin} \ +\ \hat H_{m}\ +\ \hat H_{el}\ +\ \hat H_{h}\ ,
\nonumber\\
\ & = & \
\frac{1}{2}\  
\sum_{n=0}^{2L-2}
\ \sum_{f=u,d,s,...}
\ \sum_{\alpha=r,g,b}
\left( 
\hat c_{\alpha,f,n}^\dagger \hat c_{\alpha,f,n+1}^\dagger - \hat c_{\alpha,f,n} \hat c_{\alpha,f,n+1}
\right)
\ +\ 
3  \sum_{n=0}^{2L-1}\ 
\sum_{f=u,d,s,...}
\ m_f\ \hat{\mathcal{B}}_{f,n}
\nonumber\\
&  & + \
\frac{g^2}{2}\  
\sum_{n=0}^{2L-2}\ 
\sum_{a=1}^8\ 
\left( \sum_{m\le n}  \ \sum_{f=u,d,s,...}
\hat Q^{(a)}_{f,m} \right)^2
\ +\ 
\frac{h^2}{2}\  
\sum_{n=0}^{2L-1}\ 
\sum_{a=1}^8\ 
\left( \sum_{m\le n}  \ \sum_{f=u,d,s,...}
\hat Q^{(a)}_{f,m} \right)^2
\ .
\label{eq:Hq8nfL}
\end{eqnarray}
The mapping to qu8its is a generalization of that discussed above, and displayed in Fig.~\ref{fig:2flavLayout}.
There are $2 N_f L$ qu8its, half support the quark sites and half support the anti-quark sites.
The kinetic term connects adjacent qu8its and anti-qu8its of the same flavor across the lattice, 
the mass term provides contributions from individual qu8its and anti-qu8its, and the 
chromo-electric term(s) connects all qu8its and anti-qu8its.
The kinetic term in the Hamiltonian becomes,
\begin{eqnarray}
H_{kin} &= &  
\sum_{n=0}^{2L-2}
\ \sum_{f=u,d,s,...}
\ \sum_{\alpha=r,g,b}
\frac{1}{4}(A_{0,f,n}^{(\alpha)}\otimes A_{1,f,n+1}^{(\alpha)}-B_{0,f,n}^{(\alpha)}\otimes B_{1,f,n+1}^{(\alpha)})
\ ,
\label{eq:KinnfL}
\end{eqnarray}
where $A_{0,f,n}^{(\alpha)}$ denotes $A_{0}^{(\alpha)}$
from Eq.~(\ref{eq:A12def}),
acting on the flavor $f$ qu8it at site $n$.
The mass terms given in Eq.~(\ref{eq:Hq8nfL}), with a straightforward generalization of the above, 
become
\begin{eqnarray}
H_{m} &= &  
3  \sum_{n=0}^{2L-1}\ 
\sum_{f=u,d,s,...}
\ m_f\ \widetilde{\mathcal{B}} _{f,n}
\ .
\label{eq:MnfL}
\end{eqnarray}
The color charge-charge contribution is somewhat more involved due to the number of terms, 
but can be simplified using symmetries of the sums, as shown in App.~\ref{ap:QQterms}.
The  summation can be reorganized,
\begin{eqnarray}
& & \sum_{n=0}\ 
\sum_{a=1}^8\ 
\left( \sum_{m\le n}  \ \sum_{f=u,d,s,...}
\hat Q^{(a)}_{f,m} \right)^2
\ =\  
\sum_{n=0}\ 
\sum_{m, m^\prime\le n} \ \sum_{f, f^\prime}\ 
\sum_{a=1}^8\ \hat Q^{(a)}_{f,m} \hat Q^{(a)}_{f^\prime,m^\prime} 
\nonumber\\
&  & \qquad \ =\ 
\sum_{n=0}\ 
\sum^n_{\substack{ m,m^\prime \in {\rm even}\\ m,m^\prime \in {\rm odd}}}
\sum_{f, f^\prime}\ 
\left(\sum_a \widetilde{Q}^{(a)}_{f,m} \ \widetilde{Q}^{(a)}_{f^\prime,m^\prime}\right)
\ +\ 
2 \sum_{n=0}\ 
\sum^n_{\substack{ m \in {\rm even}\\ m^\prime \in {\rm odd}}}
\sum_{f, f^\prime}\ 
\left(\sum_a \widetilde{Q}^{(a)}_{f,m}  \ \widetilde{\bar Q}{}^{(a)}_{f^\prime,m^\prime}\right)
\ ,
\label{eq:elgen}
\end{eqnarray}
where some of the terms act on the same site and flavor.
Equation~(\ref{eq:elgen}) can be expanded in terms of the $D$ and $C$ operators from Eq.~(\ref{eq:ABH1hred}).

Now that the full Hamiltonian has been 
decomposed into one-qu8it and two-qu8it operations, an estimate of quantum resources can be performed for a single Trotter step of time evolution for the qu8it mappings, and compared to previously established results for qubit mappings~\cite{Farrell:2022wyt}.
Since the Trotter decomposition used in this work is the same as the one used for qubits~\cite{Farrell:2022wyt}, the error accumulated after performing a certain amount of Trotter steps will be the same (see Fig.~14 in Ref.~\cite{Farrell:2022wyt}). The advantage of using qu8its will be in the reduction of gates required to apply a single Trotter step.
Table~\ref{tab:resources} displays the resource estimates for entangling gates for 
qu8its and qubits.
\begin{table}[!t]
    \centering
    \renewcommand{\arraystretch}{1.3}
    \begin{tabularx}{\textwidth}{C{4cm}|C{3.5cm}|C{3.5cm}|Y}
    \hline \hline
        Qudits & Number of qudits & $U_{kin}$ ent.\ gates & $U_{el}$  ent.\ gates \\\hline
        Qubit $(d=2)$ & $6 N_f L$ & $6 N_f (8L-3)-4$ & $N_f (2L-1) [23 N_f (2L-1) -17]$ \\
        Qu8it $(d=8)$ & $2 N_f L$ & $6 N_f (2L-1)$ & $4N_f (2L-1) [N_f (2L-1) -1]$ \\ \hline\hline
        Reduction in resources ($L\rightarrow \infty$) & $3$ & $4$ & $5.75$ \\\hline \hline
    \end{tabularx}
    \renewcommand{\arraystretch}{1}
    \caption{The number of qudits ($d=2$ and $d=8$) and entangling gates for applying a single Trotter step 
    using the unitary operators corresponding to the kinetic $(U_{kin})$ and the 
    ${\cal O}(g^2)$ chromo-electric $(U_{el})$ parts of the SU(3) Hamiltonian, comparing qubit and qu8it implementations.}
    \label{tab:resources}
\end{table}
The number of entangling operations required for the qu8it mapping is significantly less than for the qubit mapping (by factors $\gtrsim 5$).

\section{Summary and Outlook}
\label{sec:Summary}
\noindent
Motivated by continuing advances in the development of qudits for quantum computing, we have explored mapping 1+1D QCD to $d=8$ qudits.
We have presented the general framework  for performing quantum simulations of QCD with arbitrary number of flavors and lattice sites, and provided a detailed discussion of the theory with $N_f=1$ and $L=1$.
The main reason for considering performing quantum simulations using qu8its is because the number of two-qu8it entangling operations required to evolve a given state forward in time is significantly less 
(more than a factor of 5 reduction) than the corresponding number for mappings to qubits.
This is an important consideration for two main reasons. One is that the time to perform a two-qudit entangling operation on a quantum device is much longer than for a single qudit operation, 
and the second is the relative fidelity of the two types of operations.
The naive mapping with sequentially-Trotterized entangling operations does not provide obvious gains, but the recently developed capabilities to simultaneously induce multiple transitions within qudits, enabling multiple entangling operations to be performed in parallel, is the source of the large gain.
Thus, qudit devices of comparable fidelity gate operations and coherence times to 
an analogous device with a qubit register, 
are expected to be able to perform significantly superior quantum simulations of 1+1D QCD.

The results presented in this work readily generalize to an arbitrary number of colors.
For the $N_c=2$ case, relevant for $SU(2)$, ququarts ($d=4$) are needed to embed the vacuum in 1 state, single quarks in 2 states, and singlet two-quark in 1 state. 
The number of entangling gates for each term of the kinetic piece of the Hamiltonian is 
reduced to 4, and for each $\widetilde{Q}^{(a)}\otimes\widetilde{Q}^{(a)}$ term, 3 entangling gates are required.
For $N_c=4$, analogous gains can be achieved using qudits with $d=16$, qu16its.
The mapping is such that the vacuum occupies 1 state, single quarks occupy 4, two quarks occupy 6, three quarks occupy 4, and four quarks occupy 1. 
It requires 8 entangling gates for the kinetic piece, and 15 for each $\widetilde{Q}^{(a)}\otimes\widetilde{Q}^{(a)}$ term. 
Quarks transforming in higher-dimension gauge groups can be mapped in similar ways, with $2N_c$ terms needed for the kinetic piece, and $N_c^2-1$ for $\widetilde{Q}^{(a)}\otimes\widetilde{Q}^{(a)}$.
While the reduction in resources compared to qubits remains constant for the kinetic part, for the chromo-electric piece it is found to scale as $N_c(2N_c+17)/(3+3N_c)$, which increases as a function of $N_c$.
Mapping fermion occupations to qudits, as we have presented in this work, 
inspired by quantum chemistry and nuclear many-body systems, 
are also expected to accelerate quantum simulations of quantum field theories in 
higher numbers of spatial dimensions.
There, where both fermions and gauge bosons have to be considered, hybrid architectures might provide the most efficient mapping~\cite{Meth:2023wzd,Zache:2023cfj}.
This is the subject of future work.

\begin{acknowledgements}
\noindent
We would like to thank Crystal Senko and Sara Mouradian for discussions related 
to trapped-ion qudit systems which, in part, led to this work.
We would also like to thank Roland Farrell
for helpful discussions, and for all of our other colleagues and collaborators that provide the platform from which this work has emerged.
Martin Savage would like to thank the Physics Department at Universit\"at Bielefeld for kind hospitality during some of this work.
This work was supported, in part, by Universit\"at Bielefeld and ERC-885281-KILONOVA Advanced Grant (Caroline Robin), by U.S. Department of Energy, Office of Science, Office of Nuclear Physics, InQubator for Quantum Simulation (IQuS)\footnote{\url{https://iqus.uw.edu}} under DOE (NP) Award No.\ DE-SC0020970 via the program on Quantum Horizons: QIS Research and Innovation for Nuclear Science\footnote{\url{https://science.osti.gov/np/Research/Quantum-Information-Science}} (Martin Savage), and the Quantum Science Center (QSC),\footnote{\url{https://qscience.org}} 
a National Quantum Information Science Research Center of the U.S.\ Department of Energy (Marc Illa).
This work was supported, in part, through the Department of Physics\footnote{\url{https://phys.washington.edu}}
and the College of Arts and Sciences\footnote{\url{https://www.artsci.washington.edu}} at the University of Washington. 
We have made extensive use of Wolfram {\tt Mathematica}~\cite{Mathematica}.
\end{acknowledgements}
\clearpage
\appendix

\section{Gell-Mann Matrices}
\label{ap:GM}
\noindent
The matrix representation of the generators of SU(3) transformations, $T^a$, 
acting on the fundamental representation  are related to the 
Gell-Mann matrices via $T^a=\frac{1}{2}\lambda^a$, 
such that ${\rm Tr}[ T^a T^b ]=\frac{1}{2} \delta^{ab}$.
Using Gell-Mann's convention,
\begin{eqnarray}
\lambda^1 & = & 
\left(
\begin{array}{ccc}
0&1&0 \\ 1&0&0 \\ 0&0&0
\end{array}
\right)
\ ,\ \ 
\lambda^2 \ =\ 
\left(
\begin{array}{ccc}
0&-i&0 \\ i&0&0 \\ 0&0&0
\end{array}
\right)
\ ,\ \ 
\lambda^3 \ =\ 
\left(
\begin{array}{ccc}
1&0&0 \\ 0&-1&0 \\ 0&0&0
\end{array}
\right)
\ ,\ \ 
\lambda^4 \ =\  
\left(
\begin{array}{ccc}
0&0&1 \\ 0&0&0 \\ 1&0&0
\end{array}
\right)
\ ,
\nonumber\\
\lambda^5 & = &  
\left(
\begin{array}{ccc}
0&0&-i \\ 0&0&0 \\ i&0&0
\end{array}
\right)
\ ,\ \ 
\lambda^6 \ =\ 
\left(
\begin{array}{ccc}
0&0&0 \\ 0&0&1 \\ 0&1&0
\end{array}
\right)
\ ,\ \ 
\lambda^7 \ =\ 
\left(
\begin{array}{ccc}
0&0&0 \\ 0&0&-i \\ 0&i&0
\end{array}
\right)
\ ,\ \ 
\lambda^8 \ =\  
\frac{1}{\sqrt{3}}
\left(
\begin{array}{ccc}
1&0&0 \\ 0&1&0 \\ 0&0&-2
\end{array}
\right)
\ .
\label{eq:Murrays}
\end{eqnarray}
%

\section{Embedding Quarks and Anti-Quarks into Qu8its}
\label{ap:embedQaQ}
\noindent
The fully antisymmetric quark states can be built from the vacuum and fermionic creation operators. 
Starting  with the vacuum state,
\begin{align}
    &|1 \rangle \equiv |\Omega\rangle \ ,
\end{align}
the one-quark states are,
\begin{align}
    \hat c^\dag_r  |1 \rangle = \hat c^\dag_r|\Omega\rangle = |q_r\rangle \equiv |2\rangle \ , 
    \quad \hat c^\dag_g  |1 \rangle = \hat c^\dag_g|\Omega\rangle = |q_g\rangle \equiv |3\rangle \ , \quad \hat c^\dag_b  |1 \rangle = \hat c^\dag_b|\Omega\rangle = |q_b\rangle \equiv |4\rangle \ ,
\end{align}
the two-quark states are,
\begin{align}
    & \hat c^\dag_r  |2 \rangle = \hat c^\dag_r|q_r\rangle = 0 
    \ , 
    && \hat c^\dag_g  |2 \rangle = \hat c^\dag_g|q_r\rangle = \frac{1}{\sqrt{2}}|q_gq_r-q_rq_g\rangle \equiv -|7 \rangle 
    \ , 
    \nonumber\\
    & \hat c^\dag_r  |3 \rangle = \hat c^\dag_r|q_g\rangle = \frac{1}{\sqrt{2}}|q_rq_g-q_gq_r\rangle \equiv |7 \rangle \ , 
    && \hat c^\dag_g  |3 \rangle = \hat c^\dag_g|q_g\rangle = 0 \ , 
    \nonumber\\
    & \hat c^\dag_r  |4 \rangle = \hat c^\dag_r|q_b\rangle = \frac{1}{\sqrt{2}}|q_rq_b-q_bq_r\rangle \equiv -|6 \rangle \ , 
    && \hat c^\dag_g  |4 \rangle = \hat c^\dag_g|q_b\rangle = \frac{1}{\sqrt{2}}|q_gq_b-q_bq_g\rangle \equiv |5 \rangle \ ,  
    \nonumber\\
    & \hat c^\dag_b  |2 \rangle = \hat c^\dag_b|q_r\rangle = \frac{1}{\sqrt{2}}|q_bq_r-q_rq_b\rangle \equiv |6 \rangle \ ,
    \nonumber\\
    & \hat c^\dag_b  |3 \rangle = \hat c^\dag_b|q_g\rangle = \frac{1}{\sqrt{2}}|q_bq_g-q_gq_b\rangle \equiv -|5 \rangle \ ,
    \nonumber\\
    & \hat c^\dag_b  |4 \rangle = \hat c^\dag_b|q_b\rangle = 0 
    \ ,
\end{align}
and the  three-quark state is,
\begin{align}
    & \hat c^\dag_r  |5 \rangle = \hat c^\dag_r\frac{1}{\sqrt{2}}|q_gq_b-q_bq_g\rangle = \frac{1}{\sqrt{2}}\left(\frac{1}{\sqrt{3}}|q_rq_gq_b-q_gq_rq_b+q_gq_bq_r\rangle-\frac{1}{\sqrt{3}}|q_rq_bq_g-q_bq_rq_g+q_bq_gq_r\rangle\right) \equiv |8\rangle \ , \nonumber\\
    & \hat c^\dag_r  |6 \rangle = \hat c^\dag_r\frac{1}{\sqrt{2}}|q_bq_r-q_rq_b\rangle = 0 \ , \qquad\qquad\qquad\ \ \hat c^\dag_r  |7 \rangle = \hat c^\dag_r\frac{1}{\sqrt{2}}|q_rq_g-q_gq_r\rangle = 0 \ , \nonumber\\
    & \hat c^\dag_g  |5 \rangle = \hat c^\dag_g\frac{1}{\sqrt{2}}|q_gq_b-q_bq_g\rangle = 0 \ , \qquad\qquad\qquad\ \ \hat c^\dag_b  |5 \rangle = \hat c^\dag_b\frac{1}{\sqrt{2}}|q_gq_b-q_bq_g\rangle = 0 \ , \nonumber\\
    & \hat c^\dag_g  |6 \rangle = \hat c^\dag_g\frac{1}{\sqrt{2}}|q_bq_r-q_rq_b\rangle \equiv |8\rangle \ , \qquad\qquad\qquad \hat c^\dag_b  |6 \rangle = \hat c^\dag_b\frac{1}{\sqrt{2}}|q_bq_r-q_rq_b\rangle = 0 \ , \nonumber\\
    & \hat c^\dag_g  |7 \rangle = \hat c^\dag_g\frac{1}{\sqrt{2}}|q_rq_g-q_gq_r\rangle = 0 \ , \qquad\qquad\qquad\ \ \hat c^\dag_b  |7 \rangle = \hat c^\dag_b\frac{1}{\sqrt{2}}|q_rq_g-q_gq_r\rangle \equiv |8\rangle \ .
\end{align}
We note that with these definitions,  it is equally profitable to denote the qu8it states as
\begin{eqnarray}
    \big\{  |{\rm qu8it}\rangle  \big\}
    & = &
    \big\{ 
    |\Omega\rangle \ , \ 
    |q_r\rangle\ , \ 
    |q_g\rangle\ , \ 
    |q_b\rangle\ , \ 
    |q_g q_b\rangle\ , \ 
    -|q_r q_b\rangle\ , \ 
    |q_r q_g\rangle\ , \ 
    |q_r q_g q_b\rangle
    \big\}
    \nonumber\\
    & = &
    \big\{ 
    |1\rangle \ , \ 
    |2\rangle\ , \ 
    |3\rangle\ , \ 
    |4\rangle\ , \ 
    |5\rangle\ , \ 
    |6\rangle\ , \ 
    |7\rangle\ , \ 
    |8\rangle
    \big\}  
    \nonumber\\
    & = & 
    \big\{ 
    |\mathbf{1}_0\rangle \ , \ 
    |\mathbf{3}_1,1\rangle\ , \ 
    |\mathbf{3}_1,2\rangle\ , \ 
    |\mathbf{3}_1,3\rangle\ , \ 
    |\overline{\mathbf{3}}_2,1\rangle\ , \ 
    |\overline{\mathbf{3}}_2,2\rangle\ , \ 
    |\overline{\mathbf{3}}_2,3\rangle\ , \ 
    |\mathbf{1}_3\rangle
    \big\}
    \ ,
    \label{eq:states2_app}
\end{eqnarray}
where the subindex in the irrep labels the number of quarks in the state.
As discussed in the main text, an analogous mapping for the anti-quarks is,
\begin{eqnarray}
    \big\{  | \overline{\rm qu8it}\rangle  \big\}
    & = &
    \big\{ 
    |\phi\rangle \ , \ 
    |\overline{q}_r\rangle\ , \ 
    |\overline{q}_g\rangle\ , \ 
    |\overline{q}_b\rangle\ , \ 
    |\overline{q}_g \overline{q}_b\rangle\ , \ 
    -|\overline{q}_r \overline{q}_b\rangle\ , \ 
|\overline{q}_r \overline{q}_g\rangle\ , \ 
    |\overline{q}_r \overline{q}_g \overline{q}_b\rangle
    \big\}
    \nonumber\\
   & = &
    \big\{ 
    |\bar 1\rangle \ , \ 
    |\bar 2\rangle\ , \ 
    |\bar 3\rangle\ , \ 
    |\bar 4\rangle\ , \ 
    |\bar 5\rangle\ , \ 
    |\bar 6\rangle\ , \ 
    |\bar 7\rangle\ , \ 
    |\bar 8\rangle
    \big\}  
     \nonumber\\
    & = & 
    \big\{ 
    |\mathbf{1}_0\rangle \ , \ 
    |\overline{\mathbf{3}}_1,1\rangle\ , \ 
    |\overline{\mathbf{3}}_1,2\rangle\ , \ 
    |\overline{\mathbf{3}}_1,3\rangle\ , \ 
    |\mathbf{3}_2,1\rangle\ , \ 
    |\mathbf{3}_2,2\rangle\ , \ 
    |\mathbf{3}_2,3\rangle\ , \ 
    |\mathbf{1}_3\rangle
    \big\}    
    \ \ .
\end{eqnarray}
%

\section{Givens Rotations for SU(8)}
\label{ap:su8givens}
\noindent
Givens rotations are a straightforward way to access SU(8) transformations, and 
are particularly convenient for quantum operations that are sequential applications of two-level transformations, 
as are induced, for example, by application of lasers to trapped ions.
The notation that we will use parallels and extends the notation used for Pauli operators,
$\sigma_x$, $\sigma_y$ and $\sigma_z$ $\rightarrow$ ${\cal X}_{ij}$, ${\cal Y}_{ij}$ and ${\cal Z}_{i}$.
For SU(8) transformations, there are 28 ${\cal X}_{ij}$s, 28 ${\cal Y}_{ij}$s, and 7 ${\cal Z}_{i}$s.

For the 7 diagonal generators ${\cal Z}_{i}$s, one basis is a straightforward extension of  Gell-Mann's SU(3): 
\begin{eqnarray}
{\cal Z}_{1} \ & = & \ {\rm diag}\left(1,-1,0,0,0,0,0,0\right)
\ ,\ \
{\cal Z}_{2} \ = \ \frac{1}{\sqrt{3}} {\rm diag}\left(1,1,-2,0,0,0,0,0\right)  \ ,
\nonumber \\
{\cal Z}_{3} \ & = & \ \frac{1}{\sqrt{6}} {\rm diag}\left(1,1,1,-3,0,0,0,0\right)
\ ,\ \
{\cal Z}_{4} \ =\ \frac{1}{\sqrt{10}} {\rm diag}\left(1,1,1,1,-4,0,0,0\right) \ ,
\nonumber \\
{\cal Z}_{5} \ & = & \ \frac{1}{\sqrt{15}} {\rm diag}\left(1,1,1,1,1,-5,0,0\right) 
\ ,\ \
{\cal Z}_{6} \ = \ \frac{1}{\sqrt{21}} {\rm diag}\left(1,1,1,1,1,1,-6,0\right) \ ,
\nonumber \\
{\cal Z}_{7} \ & = & \ \frac{1}{\sqrt{28}} {\rm diag}\left(1,1,1,1,1,1,1,-7\right) 
\ ,\ \ 
{\rm Tr} \left[{\cal Z}_{i} {\cal Z}_{j} \right] \ = \  2 \delta^{ij}
\ .
\end{eqnarray}
However, an alternate choice that makes better connection to sequency analysis, 
and which we have chosen to use in our work, 
is the Hadamard-Walsh basis, 
which includes the identity operator, 
\begin{eqnarray}
w_1 \ & = & \ \frac{1}{\sqrt{8}} {\rm diag}\left( 1,1,1,1,1,1,1,1 \right)
\ ,\ \ 
w_2 \ = \  \frac{1}{\sqrt{8}} {\rm diag}\left( 1,1,1,1,-1,-1,-1,-1 \right) \ ,
\nonumber\\
w_3 \ & = & \ \frac{1}{\sqrt{8}} {\rm diag}\left( 1,1,-1,-1,-1,-1,1,1 \right)
\ ,\ \ 
w_4 \ = \  \frac{1}{\sqrt{8}} {\rm diag}\left( 1,1,-1,-1,1,1,-1,-1 \right) \ ,
\nonumber\\
w_5 \ & = & \ \frac{1}{\sqrt{8}} {\rm diag}\left( 1,-1,-1,1,1,-1,-1,1 \right)
\ ,\ \ 
w_6 \ = \  \frac{1}{\sqrt{8}} {\rm diag}\left( 1,-1,-1,1,-1,1,1,-1 \right) \ ,
\nonumber\\
w_7 \ & = & \ \frac{1}{\sqrt{8}} {\rm diag}\left( 1,-1,1,-1,-1,1,-1,1 \right)
\ ,\ \ 
w_8 \ = \  \frac{1}{\sqrt{8}} {\rm diag}\left( 1,-1,1,-1,1,-1,1,-1 \right)
\ ,
\end{eqnarray}
which are normalized such that 
${\rm Tr} \left[ w_{i}  w_{j} \right]  =  \delta^{ij}$.
The ordering of the $w_i$ is the same as those produced 
in {\tt Mathematica} from {\tt HadamardMatrix[8]}.

The matrix representations of the (symmetric) $\sigma^x$-type generators, 
${\cal X}_{ij}$, have all zero entries except for the elements defined by 
${ij}$, such that ${\cal X}_{ij}={\cal X}_{ji}=1$.
Similarly, the matrix representations of the (anti-symmetric) $\sigma^y$-type generators, 
${\cal Y}_{ij}$, have all zero entries except for the elements defined by 
${ij}$, such that ${\cal Y}_{ij}=-{\cal Y}_{ji}=-i$.
Examples are:
\begin{eqnarray}
{\cal X}_{13} & = & 
\left(
\begin{array}{cccccccc}
0 & 0 & 1 & 0 & 0 & 0 & 0 & 0 \\
0 & 0 & 0 & 0 & 0 & 0 & 0 & 0 \\
1 & 0 & 0 & 0 & 0 & 0 & 0 & 0 \\
0 & 0 & 0 & 0 & 0 & 0 & 0 & 0 \\
0 & 0 & 0 & 0 & 0 & 0 & 0 & 0 \\
0 & 0 & 0 & 0 & 0 & 0 & 0 & 0 \\
0 & 0 & 0 & 0 & 0 & 0 & 0 & 0 \\
0 & 0 & 0 & 0 & 0 & 0 & 0 & 0 
\end{array}
\right)
\ , \ \
{\cal Y}_{13} \ =\  
\left(
\begin{array}{cccccccc}
0 & 0 & -i & 0 & 0 & 0 & 0 & 0 \\
0 & 0 & 0 & 0 & 0 & 0 & 0 & 0 \\
i & 0 & 0 & 0 & 0 & 0 & 0 & 0 \\
0 & 0 & 0 & 0 & 0 & 0 & 0 & 0 \\
0 & 0 & 0 & 0 & 0 & 0 & 0 & 0 \\
0 & 0 & 0 & 0 & 0 & 0 & 0 & 0 \\
0 & 0 & 0 & 0 & 0 & 0 & 0 & 0 \\
0 & 0 & 0 & 0 & 0 & 0 & 0 & 0 
\end{array}
\right)
\ .
\label{eq:XYdef}
\end{eqnarray}
%

\section{Contractions of Color-Charge Operators}
\label{ap:QQterms}
\noindent
Expressions for the contractions of color charge-charge operators can be found straightforwardly.
Acting on a qu8it lattice site twice (for a 2-site system), 
the summations over adjoint indices reduce to 
\begin{eqnarray}
\sum_a\ \left( \widetilde{Q}^{(a)}\otimes \widetilde{I} \right)^2 
& = & \
\frac{4}{3} 
{\rm diag}\left( 0,1,1,1,1,1,1,0 \right) \otimes \widetilde{I}
\nonumber\\
& = & \
8 w_1\otimes w_1 - \frac{8}{3}\left( w_3 + w_5 + w_7 \right) \otimes w_1 
\nonumber\\
& = & \
\widetilde{I}\otimes \widetilde{I} - {\rm diag}\left(1,-\frac{1}{3},-\frac{1}{3},-\frac{1}{3},-\frac{1}{3},-\frac{1}{3},-\frac{1}{3},1\right)\otimes \widetilde{I}
\ ,
\end{eqnarray}
and acting on a anti-qu8it lattice site twice, 
the summations over adjoint indices reduce to 
\begin{eqnarray}
\sum_a\ \left( \widetilde{I} \otimes \widetilde{\bar Q}{}^{(a)}  \right)^2 
& = & \
\frac{4}{3} 
\widetilde{I} \otimes  {\rm diag}\left( 0,1,1,1,1,1,1,0 \right)  
\nonumber\\
& = & \
8 w_1\otimes w_1 - \frac{8}{3} w_1 \otimes \left( w_3 + w_5  + w_7 \right) 
\nonumber\\
& = & \
\widetilde{I} \otimes \widetilde{I} - 
\widetilde{I} \otimes {\rm diag}\left(1,-\frac{1}{3},-\frac{1}{3},-\frac{1}{3},-\frac{1}{3},-\frac{1}{3},-\frac{1}{3},1\right)
\ .
\end{eqnarray}

In general, there will be contributions from color-charge operators acting on two sites,
of the form quark-quark, anti-quark-anti-quark and quark-anti-quark.
For color-charge operators acting on arbitrary quark-quark 
and anti-quark-anti-quark sites, 
\begin{eqnarray}
\sum_a\  
\widetilde{Q}^{(a)}\otimes \widetilde{Q}^{(a)} 
\ =\ 
\sum_a \widetilde{\bar Q}{}^{(a)}\otimes \widetilde{\bar Q}{}^{(a)} 
\ & = & \
\frac{1}{2}\left(w_3-w_5\right)\otimes \left(w_3-w_5\right)
\ +\ \frac{1}{6} 
\left(w_4+w_6-2 w_8\right) \otimes \left(w_4+w_6-2 w_8\right)
\nonumber\\
&  & + \ 
\frac{1}{4} 
\sum_{\substack{ r \in \{(23),(24),(34), \\ 
\ \ (56),(57),(67)\}}}
\left( {\cal Y}_r\otimes {\cal Y}_r + {\cal X}_r\otimes {\cal X}_r  \right)
\nonumber\\
& & + \
\frac{1}{4} 
\sum_{\substack{ (r,s) \in \{(23)(56), \\ \ \ \ \ 
(24)(57), (34)(67)\}}}
\left(
{\cal Y}_r\otimes {\cal Y}_s - {\cal X}_r\otimes {\cal X}_s 
+ {\cal Y}_s\otimes {\cal Y}_r - {\cal X}_s\otimes {\cal X}_r 
\right)
\ ,\
\end{eqnarray}
with
\begin{eqnarray}
w_3-w_5 & = & \frac{1}{\sqrt{2}} {\rm diag}\left( 0,1,0,-1,-1,0,1,0\right)
\ ,\ 
w_4+w_6-2 w_8 \ =\  \frac{1}{\sqrt{2}} {\rm diag} \left(0,1,-2,1,-1,2,-1,0 \right)
\ .
\end{eqnarray}

Similarly, for operators acting on quark-anti-quark or anti-quark-quark sites, 
\begin{eqnarray}
\sum_a\  
\widetilde{Q}^{(a)} \otimes 
\widetilde{\bar Q}{}^{(a)}
\ =\ 
\sum_a\  
\widetilde{\bar Q}{}^{(a)}\otimes \widetilde{Q}^{(a)} 
& = & 
- \ \frac{1}{2}\left(w_3-w_5\right)\otimes \left(w_3-w_5\right)
\ -\ \frac{1}{6} 
\left(w_4+w_6-2 w_8\right) \otimes \left(w_4+w_6-2 w_8\right)
\nonumber\\
& & + \
\frac{1}{4} 
\sum_{\substack{ r \in \{(23),(24),(34), \\ 
\ \ (56),(57),(67)\}}}
\left(  {\cal Y}_r\otimes {\cal Y}_r - {\cal X}_r\otimes {\cal X}_r \right)
\nonumber\\
& & + \
\frac{1}{4} 
\sum_{\substack{ (r,s) \in  \{(23)(56), \\ \ \ \ \ 
(24)(57), (34)(67)\}}}
\left(
{\cal X}_r\otimes {\cal X}_s + {\cal Y}_r\otimes {\cal Y}_s
+
{\cal X}_s\otimes {\cal X}_r + {\cal Y}_s\otimes {\cal Y}_r
\right)
\ \ .
\end{eqnarray}

Rewriting these sums in terms of commuting operators gives,
\begin{eqnarray}
\sum_a \widetilde{Q}^{(a)}\otimes \widetilde{Q}^{(a)}
& = & 
\frac{1}{4}(  D^{(12)} \otimes D^{(12)} + C^{(12)} \otimes C^{(12)})
    +\frac{1}{4}(  D^{(45)} \otimes D^{(45)} + C^{(45)} \otimes C^{(45)}) 
    \nonumber \\ 
    & &  
    + \ \frac{1}{4}(  D^{(67)} \otimes D^{(67)} + C^{(67)} \otimes C^{(67)} )
\ +\ \widetilde{Q}^{(3)}\otimes \widetilde{Q}^{(3)}
\ +\ \widetilde{Q}^{(8)}\otimes \widetilde{Q}^{(8)} \ ,
    \label{eq:elregroupAPP}
\end{eqnarray}
\begin{eqnarray}
\sum_a \widetilde{\bar Q}{}^{(a)}\otimes \widetilde{\bar Q}{}^{(a)}
& = & 
\frac{1}{4}(  D^{(12)} \otimes D^{(12)} + C^{(12)} \otimes C^{(12)})
    +\frac{1}{4}(  D^{(45)} \otimes D^{(45)} + C^{(45)} \otimes C^{(45)}) 
    \nonumber \\ 
    & &  
    + \ \frac{1}{4}(  D^{(67)} \otimes D^{(67)} + C^{(67)} \otimes C^{(67)} )
\ +\ \widetilde{\bar Q}{}^{(3)}\otimes \widetilde{\bar Q}{}^{(3)}
\ +\ \widetilde{\bar Q}{}^{(8)}\otimes \widetilde{\bar Q}{}^{(8)}
\nonumber\\
& = & \sum_a \widetilde{Q}^{(a)}\otimes \widetilde{Q}^{(a)} \ ,
    \label{eq:elregroupAPPb}
\end{eqnarray}
\begin{eqnarray}
\sum_a \widetilde{Q}^{(a)}\otimes \widetilde{\bar Q}{}^{(a)}
& = & 
\frac{1}{4}(  D^{(12)} \otimes D^{(12)} - C^{(12)} \otimes C^{(12)})
    +\frac{1}{4}(  D^{(45)} \otimes D^{(45)} - C^{(45)} \otimes C^{(45)}) 
    \nonumber \\ 
    & &  
    + \ \frac{1}{4}(  D^{(67)} \otimes D^{(67)} - C^{(67)} \otimes C^{(67)} )
\ +\ \widetilde{Q}^{(3)}\otimes \widetilde{\bar Q}{}^{(3)}
\ +\ \widetilde{Q}^{(8)}\otimes \widetilde{\bar Q}{}^{(8)} \ ,
    \label{eq:elregroupAPPc}
\end{eqnarray}
\begin{eqnarray}
\sum_a \widetilde{\bar Q}{}^{(a)}\otimes \widetilde{Q}^{(a)}
& = & 
\frac{1}{4}(  D^{(12)} \otimes D^{(12)} - C^{(12)} \otimes C^{(12)})
    +\frac{1}{4}(  D^{(45)} \otimes D^{(45)} - C^{(45)} \otimes C^{(45)}) 
    \nonumber \\ 
    & &  
    + \ \frac{1}{4}(  D^{(67)} \otimes D^{(67)} - C^{(67)} \otimes C^{(67)} )
\ +\ \widetilde{\bar Q}{}^{(3)}\otimes \widetilde{Q}^{(3)}
\ +\ \widetilde{\bar Q}{}^{(8)}\otimes \widetilde{Q}^{(8)}
\nonumber\\
& = & \sum_a \widetilde{Q}^{(a)}\otimes \widetilde{\bar Q}{}^{(a)}
\ ,
\label{eq:elregroupAPPd}
\end{eqnarray}
where the $C^{(ij)}$ and $D^{(ij)}$ are given in Eq.~(\ref{eq:ABH1hred}),
and 
$\widetilde{\bar Q}{}^{(3)} = - \widetilde{Q}^{(3)}$ and 
$\widetilde{\bar Q}{}^{(8)} = - \widetilde{Q}^{(8)}$
have been used.

\FloatBarrier
\bibliography{bibi_qu8its}

\end{document}